# Tara Polaris expeditions: Sustained decadal observations of the coupled Arctic system in rapid transition

**Authors**
Mathieu Ardyna[1,2]* (Mathieu.Ardyna@takuvik.ulaval.ca), Marcel Nicolaus[3],* (Marcel.Nicolaus@awi.de), Marie-Noëlle Houssais[4] (marie-noelle.houssais@locean.ipsl.fr), Jean-Christophe Raut[4,5] (jean-christophe.raut@latmos.ipsl.fr), Hélène Angot[6] (helene.angot@univ-grenoble-alpes.fr), Kelsey Bisson[7,8] (kelsey.bisson@nasa.gov), Kristina A. Brown[9] (Kristina.Brown@umanitoba.ca), J. Michel Flores[10] (flores@weizmann.ac.il), Pierre E. Galand[11] (pierre.galand@obs-banyuls.fr), Jean-François Ghiglione[12] (ghiglione@obs-banyuls.fr), Maxime Geoffroy[13,14,15] (maxime.geoffroy@mi.mun.ca), Lars-Eric Heimbürger-Boavida[16] (lars-eric.heimburger@mio.osupytheas.fr), Kathy S. Law[5] (Kathy.Law@latmos.ipsl.fr), Connie Lovejoy[1] (connie.lovejoy@bio.ulaval.ca), François Ravetta[5] (francois.ravetta@latmos.ipsl.fr), Søren Rysgaard[17] (rysgaard@au.dk), Julia Schmale[18] (julia.schmale@epfl.ch), Nina Schuback[19] (nina.schuback@swisspolar.ch), Jeroen E. Sonke[20] (jeroen.sonke@get.omp.eu), Martin Vancoppenolle[4] (martin.vancoppenolle@locean.ipsl.fr), Jean-Eric Tremblay[1] (Jean-Eric.Tremblay@bio.ulaval.ca), Marcel Babin[1,2] (marcel.babin@takuvik.ulaval.ca), Chris Bowler[2,21] (cbowler@biologie.ens.fr), Lee Karp-Boss[22] (lee.karp-boss@maine.edu), Romain Troublé[23] (romain@fondationtaraocean.org).

Corresponding authors: Mathieu Ardyna* (Mathieu.Ardyna@takuvik.ulaval.ca), Marcel Nicolaus* (Marcel.Nicolaus@awi.de)

*The first two authors contributed equally

**Affiliations**
[1]Takuvik International Research Laboratory (IRL 3376), ULaval - CNRS - Sorbonne Université, Université Laval, Quebec, Canada.
[2]Research Federation for the study of Global Ocean Systems Ecology and Evolution, FR2022/Tara Oceans GOSEE, Paris, France
[3]Alfred-Wegener-Institut Helmholtz-Zentrum für Polar- und Meeresforschung, Bremerhaven, Germany
[4]Laboratoire d'Océanographie et du Climat : Expérimentations et Approches Numériques, Sorbonne Université, CNRS, IRD, MNHN, Paris, France
[5]LAboratoire ATMosphères Observations Spatiales, Sorbonne Université, UVSQ, CNRS, Paris, France
[6]Univ. Grenoble Alpes, CNRS, INRAE, IRD, Grenoble INP, IGE, Grenoble, France
[7]Ocean Biology and Biogeochemistry program, Earth Science Division, Science Mission Directorate, NASA Headquarters, Washington, DC, United States
[8]Department of Botany and Plant Pathology, Oregon State University, Corvallis, OR, United States
[9]Department of Environment and Geography, Centre for Earth Observation Science, University of Manitoba, Manitoba, Canada
[10]Weizmann Institute of Science, Department of Earth and Planetary Science, Rehovot, Israel.

[11]Sorbonne Université, CNRS, Laboratoire d'Ecogéochimie des Environnements Benthiques (LECOB), Observatoire Océanologique de Banyuls, Banyuls sur Mer, France
[12]Laboratoire d'Océanographie Microbienne, LOMIC, UMR7621, CNRS Sorbonne Université.
[13]Centre for Fisheries Ecosystems Research, Fisheries and Marine Institute of Memorial
[14]University of Newfoundland and Labrador, St. John's, Newfoundland and Labrador, Canada
[15]Department of Arctic and Marine Biology, UiT The Arctic University of Norway, Tromsø, Norway
[16]Aix Marseille Université, CNRS/INSU, Université de Toulon, IRD, Mediterranean Institute of Oceanography (MIO), Marseille, France
[17]CIFAR, Department of Biology, Aarhus University, 8000 Aarhus, Denmark
[18]Extreme Environments Research Laboratory, École Polytechnique Fédérale de Lausanne, EPFL Valais Wallis, Sion, Switzerland
[19]Swiss Polar Institute, Sion, Switzerland
[20]Géosciences Environnement Toulouse, CNRS/IRD/Université de Toulouse, 31400 Toulouse, France.
[21]Institut de Biologie de l'Ecole Normale Supérieure (IBENS), Ecole Normale Supérieure
[22]School of Marine Sciences /Climate Change Institute, University of Maine, Orono, ME 04473.
[23]Tara Ocean Foundation, France.

# Abstract

The coupled Arctic system is in rapid transition and is set to undergo further dramatic changes over the coming decades. These changes will lead most likely to an ice-free ocean in summer, expected before mid-century. The Arctic will become more strongly influenced by atmospheric and oceanographic processes characteristic of mid-latitudes, increasing the prevalence of contaminants and new biological species. This ongoing transition of the Arctic to a new state necessitates systematic monitoring of all sentinels (variables that make an essential contribution to characterizing the Earth's state) to improve our understanding of the system, enhance forecasting and support knowledge-based decisions. Here, we describe a sustained multi-decadal observation program to be implemented on the Tara Polar Station between 2026 and 2046. The monitoring program is designed as a series of year-long drift expeditions, called Tara Polaris, in the central Arctic Ocean, covering all seasons. The multidisciplinary data will bridge ecological, geochemical, biological, and physical parameters and processes in the atmosphere, sea ice and ocean. In addition, data collected with consistent methodologies over a 20-year period will make it possible to distinguish long-term trends from seasonal and interannual variability. In this paper, we discuss specific measurement challenges in each compartment (i.e., atmosphere, sea ice and ocean) along key sentinels and the most pressing scientific questions to be addressed. The expected outcomes of the Tara Polaris program will enable us to understand and quantify the main feedbacks of the coupled Arctic system, with their seasonal and interannual trends and spatial variability.

# 1. Introduction and state of the art

Over the past few decades, the Arctic has been warming more than twice as fast as the rest of the planet due to Arctic amplification. This rapid warming trend continues, and simulations indicate that warming has recently increased more than fourfold in some Arctic regions compared with the global average (Coupled Model Intercomparison Project, CMIP6; Rantanen et al., 2022). In addition, the expected increase in Arctic precipitation is higher in CMIP6 than previously estimated in CMIP5, due to more warming globally, enhanced poleward moisture transport and heat intrusions (Dada et al., 2022), and generally a greater sensitivity of precipitation to Arctic warming. The shift from a snow-dominated Arctic to a rain-dominated Arctic in summer and autumn could occur decades earlier than previously expected, at a lower global warming threshold, potentially below 1.5°C, with profound climatic, ecosystem and socio-economic impacts (McCrystall et al., 2021).

The decline of the Arctic sea-ice cover is one of the most striking changes in the global climate system. Summer sea-ice extent has decreased >10% per decade since the beginning of satellite observations in the late 1970s. The properties of this ice have also changed dramatically, towards a thinner, younger and more dynamic ice pack (Kwok, 2018; Sumata et al., 2023). The ongoing changes of the ice pack in all seasons over the coming 20 years are expected to result in perturbations in the seasonality of the snow cover on sea ice (Webster et al., 2014), and the distribution of summer melt ponds and melt water (Niehaus et al., 2023; Smith et al., 2023). Climate projections predict an ice-free late summer Arctic Ocean between 2030 and 2050 (Notz and SIMIP community, 2020; Jahn et al., 2024), implying the complete loss of perennial sea ice, with increasingly long open water seasons (Lebrun et al., 2019). Expected consequences are, among others, increased surface warming due to lower surface albedo and higher absorption of solar energy (Pistone et al., 2019), increased energy, momentum, and gas exchange between the atmosphere and ocean, and the loss of sea ice as a habitat for mammals, fish, amphipods, algae and microorganisms (Ardyna and Arrigo, 2020; Flores et al., 2023). Social-economic consequences may include potentially increased shipping (goods and tourism) and fishery activities, other commercial opportunities and changes in geopolitical frameworks.

The declining Arctic sea ice interacts with the rest of the climate system through numerous feedback processes (Goosse et al., 2018), notably in the Arctic atmosphere. Less ice leads to more solar short-wave absorption into the surface ocean; at the same time the warmer surface means more long-wave radiation emission to the atmosphere (Pithan and Mauritsen, 2014). Changes in the Arctic also lead to changes in poleward energy transport. Alterations in boundary layer stratification, as well as the impact of sea-ice loss on clouds and their vertical distribution are also expected. Although increased turbulent heat and moisture transport resulting from prolonged open ocean periods fosters low-cloud formation (Kay and Gettelman, 2009), a decrease in low-level clouds and a simultaneous rise in middle-level cloud cover due to decreased static stability and a deepening atmospheric boundary layer may result from ocean heating (Schweiger et al., 2008; Porter et al., 2022). An increased cloud cover contributes, in turn, to enhanced downward longwave radiation (Maillard et al., 2021), increasing the surface air temperature and prolonging the sea-ice melt season, thus forming a positive feedback resulting in additional Arctic sea-ice loss (Serreze and Barry, 2011). The lifetime of Arctic clouds and their interactions with precipitation are also significantly influenced by their interactions with aerosols (Zieger et al., 2023; Heslin-Rees et al., 2024). Arctic aerosol types and concentration are characterized by marked temporal and vertical variability, originating from diverse sources (Schmale et al., 2022; Mölders and Friberg, 2023). These aerosols possess chemical and physical properties that determine their impacts on radiation and clouds, influenced by source regions ranging from locally produced marine biogenic particles (Freitas et al., 2023) to long-range transport of natural and anthropogenic aerosols and precursors (Raut et al., 2017; Ansmann et al., 2023; Boyer et al., 2023). Declining Arctic sea ice cover also allows more $CO_2$ exchange across the air-sea interface, contributing to enhance the Arctic Ocean $CO_2$ sink (Yasunaka et al., 2023).

Marked hydrodynamical changes have been reported in the central Arctic Ocean (CAO). The maintenance of the colder and fresher surface layer is essential for maintaining the Arctic Ocean's ice cover (e.g., Toole et al., 2010). The upper ocean dynamics depend on a subtle balance of exchanges with the atmosphere and sea ice, and the large freshwater discharge from the Arctic rivers. The amount of freshwater stored in the Arctic Ocean increased by 30% in the period 1992–2012, associated with overall decreasing upper ocean salinities (Rabe et al., 2014; Haine et al., 2015) and a major contribution from freshwater accumulation in the Beaufort Gyre (Proshutinsky et al., 2020). In parallel, strong upper ocean salinification has been observed in the Eurasian Basin, associated with weakened stratification (Polyakov et al., 2020a), along with widespread upper Arctic Ocean warming. Increased solar radiation input has warmed the surface layer and contributed to increased sea-ice melt, particularly in the marginal ice zone (Steele et al., 2010; Carmack et al., 2015). Successive warm pulses in the Atlantic Water (AW) inflow at the entrance of the Arctic Ocean (Beszczynska-Möller et al., 2012; Muilwijk et al., 2018) have led to a warming of the AW layer in the Eurasian Basin and possibly contributed to regional sea-ice decline (Polyakov et al., 2010; 2020b). Further expansion of this layer in the basin may help maintain a seasonally ice-free Arctic Ocean in the future (Årthun et al., 2019). Yet, mixing of the AW heat from the ocean boundaries into the interior basins as well as its vertical spreading depend on a variety of mechanisms and forcing, including the role of mesoscale eddies, which are difficult to assess (Lenn et al., 2022). Ice-ocean interactions are constrained by, and impact, Arctic Ocean dynamics at multiple scales (Wang et al., 2019; Gupta et al., 2020), with a strong dependence on the atmospheric forcing modes of variability (Morison et al., 2021; Polyakov et al., 2023; Lin et al., 2023) and basin-scale dynamical balances (Spall, 2020; Meneghello et al., 2021).

Arctic marine ecosystems are known for a wide diversity of more than 5,000 animal species (including marine mammals and birds) and over 2,000 species of algae and other protists (AMAP, 2021), thus contributing directly to global biodiversity (Ibarbalz et al., 2023). Arctic sea-ice ecosystems are home to endemic species spanning multiple scales and trophic functions, from micrometers, with specialized microbial and algal communities, to millimeter-sized zooplankton to meters, with top predators such as the polar bear (*Ursus maritimus*). With the sea-ice thinning and disappearance, some Arctic species may be able to adapt their physiology and phenology, or move northward, in search of more favorable conditions, but others may be excluded as the Arctic warms. Current trends indicate that species that depend on sea ice for breeding, resting, or feeding will see their population range shrink as sea ice retreats earlier and the open-water season lengthens. Arctic marine species and ecosystems are therefore under pressure from cumulative changes in their physical, chemical, and biological environment (Arrigo et al., 2020). Some changes may be gradual over years or decades, but there can also be sudden, major changes that can affect ecosystem structure and functioning (AMAP, 2021). In addition, contagious diseases, pathogens and harmful algal blooms are becoming increasingly common (Waits et al., 2018; Anderson et al., 2022; Rode et al., 2024).

The Arctic Ocean is strongly impacted by wildfire activities and anthropogenic inputs (e.g., soot, mercury, plastics, organic and emerging contaminants). Contaminants are transported over long distances via rivers, oceans and air. They can have considerable negative effects on the environment, biota and human health (AMAP, 2021). Mercury contamination in the Arctic also remains a major concern, particularly in the organic form of methylmercury, which bioaccumulates and biomagnifies to harmful levels in Arctic biota (Dietz et al., 2022). Plastics (from nano- to macro-plastics) are present in all Arctic compartments (Kanhai et al., 2020, Materić et al., 2022) and can be ingested by mammals, seabirds, fish and invertebrates (Bergmann et al., 2022). Inside organisms, microplastics release toxic chemicals such as additives and monomers that disrupt endocrine function and increase risk for neurodevelopmental disorders, reproductive birth defects, infertility, cardiovascular disease, and cancers (Landrigan et al., 2023). New legacy pollutants termed 'forever chemicals' are an emerging problem in the Arctic Ocean, as they persist in the environment longer than other man-made substances (AMAP, 2021). Some pollutants are produced in the Arctic, for example, by fossil fuel combustion, and increasing industrial activities will further increase these local contributions (Law et al., 2017; Raut et al., 2022). Shipping and tourism are expanding rapidly, raising concerns about environmental impacts on Arctic ecosystems and communities. The number of ships entering the Arctic Polar Code area between 2013 and 2023 increased by 37%. The summer tourism quadrupled and winter tourism increased by over 600% between 2006 and 2016, although large areas of the Arctic remain untouched by tourism (Runge et al., 2020).

To improve understanding of the Arctic Ocean and ongoing climate- and anthropogenic-induced changes, we describe here how Tara Polar Station (TPS) will be used as a platform for the observations needed to implement a multi-decadal monitoring program in the CAO over all seasons, while drifting with the sea ice (Figure 1). A unique and innovative approach is needed, combining targeted scientific observations and long-term monitoring. The expected observations, using state-of-the-art instrumentation, will enable us to study processes as well as the spatial and temporal distribution and evolution of key sentinels in the ocean, sea ice and atmosphere. Importantly, these observations will discriminate between long-term trends and internal and year-to-year variability. Capturing seasonality is key, but also extremely challenging due to the very limited and difficult access to the CAO in winter. The following sections describe the objectives (Section 2), define the sentinels (Section 3), and describe the approaches (Section 4) and expected outcomes (Section 5). Publications describing the platform itself, and more targeted studies of sub-systems of the atmosphere, the sea ice, the ocean, and contaminants are ongoing.

## 2. Objectives and scientific questions

The TPS, with a size of 26 m x 16 m and maximum crew of 16 technical and scientific personnel will have a much smaller footprint on the surrounding environment compared to conventional research icebreakers. To bridge larger spatial scales (up to the entire CAO) and seasonal coverage in a multi-year context, TPS data will be linked to remote sensing (aircraft and satellite) observations, combined with re-analyses data, and incorporated into multi-scale numerical model simulations. The perspective of the proposed 20-year Tara Polaris program is 'long term' compared to conventional research projects and can be compared with ambitious era projects such as 41 Russian North Pole drifting ice camps (1937–2015, revived in 2022; Frolov et al., 2005). Other previous drifting efforts in the CAO have been short-term and more sporadic but generally highly interdisciplinary, such as the Alpha ice station (1957–1958; Untersteiner, 1961), SHEBA (1997–1998; Perovich et al., 1999), the drifting schooner Tara Arctic Damoclès (2006–2008; Gascard et al., 2008), N-ICE (2015; Granskog et al., 2018), IAOOS (2013–2019; Maillard et al., 2021) and North Pole-41 (2022–2024). The most comprehensive study to date in the CAO, focusing on the annual cycle of the coupled Arctic system has been the MOSAiC expedition (2019–2020; Shupe et al., 2022; Nicolaus et al., 2022; Rabe et al., 2022; Fong et al., 2024). The Tara Polaris program will build on these previous campaigns and further benefit from the existing comprehensive network of autonomous data acquisition systems (e.g., Rabe et al., 2024) and future more conventional expeditions across the Arctic Ocean.

The main objective of the Tara Polaris interdisciplinary and multi-decadal observational program is to quantify and increase understanding of the state of, and linkages between, the atmosphere, sea ice, and ocean in the CAO. Based on this improved understanding of processes and parameters, the aim is to evaluate variability and changes across different scales in time and space. The output will be based on systematic observations of the main sentinels (Figure 2; Tables 1 and 2) over the coming decades. We define a sentinel here as a physical, chemical or biological variable or group of related variables that make a critical contribution to the characterization of global change in analogy to the 'essential climate variables' as defined by the Global Climate Observing System. These sentinels will provide the empirical evidence needed to mechanistically understand and predict the effect of global change, to guide mitigation and adaptation measures, to assess risk and enable the attribution of trends in sentinels to underlying causes, and to underpin climate services. Sentinels are identified on the basis of their: 1) relevance – the variable is essential for characterizing the CAO and its changes; 2) feasibility – observation or derivation of the variable on a large scale is technically feasible using proven and scientifically understood methods; and 3) cost-effectiveness – production and archiving of data for the given variable is affordable and relies primarily on coordinated observing systems using proven technologies, drawing on historical data sets wherever possible.

Covering physical processes, biogeochemical cycles, and the composition and functioning of communities of living organisms in parallel is essential. With the initial phase of Tara Polaris drifts, together with observations from process-oriented projects, the long-term aspect will enable us to quantify ongoing and potential future changes. The following research questions specifically address what can likely be achieved with the approach:

- Question 1: What are the seasonal and interannual trends and spatial variability (from pack ice to basin scale) of the sentinels?
- Question 2: What are the main feedbacks in the coupled atmosphere-sea-ice-ocean system of the Arctic, and how are they changing in response to climate change?
- Question 3: How can the long-term and high-resolution observations from TPS contribute to the detection and quantification of the Arctic's transition toward a new state?

TPS drifts will enable the study and quantification of long-term processes covering at least two decades. In addition, they will provide a unique opportunity to assess how organisms adapt to changing environmental conditions and to quantify the impacts of pollution across the atmosphere, the snow and sea ice, and the ocean compartments. Together, these observations will help to elucidate adaptive functions and pollution responses as integral parts of the coupled Arctic system. Here, we describe how these three overarching questions relate to individual key sentinels.

To ensure the robustness of the Tara Polaris observing strategy in the harsh and dynamic conditions of the CAO, the TPS is designed with operational flexibility and redundancy. While a large fraction of the core measurements rely on installations outside the vessel (e.g., surface towers, under-ice platforms, ice-based operations), the TPS also integrates a suite of onboard sampling and observational capabilities to ensure continuous data acquisition. This suite includes a moonpool for protected under-ice access, an in-line seawater flow-through system for continuous surface ocean observations, a meteorological mast, and laboratory space for automated and manual sampling. When weather conditions or floe instability prevent outdoor operations, these internal facilities allow continued key atmospheric, sea-ice, and oceanic observations. Contingency protocols are in place to adapt sampling operations during transitions between ice-covered and open-water phases, ensuring that critical sentinels can still be monitored. A technical overview of the TPS infrastructure and operational modes will be provided in a companion paper (C Moulin, personal communication).

## 3. Categories of long-term sentinels

Establishing a reliable set of long-term sentinels is essential for creating a coordinated Arctic observation strategy. These sentinels fall into three main system components: atmosphere, sea ice and snow, and ocean; pollutant sentinels distribute across these system components (Figure 2). Each component is associated with specific parameters that collectively provide a comprehensive understanding of the coupled Arctic system. This framework serves as a starting point for implementing coordinated long-term observations during the Tara Polaris expeditions. Importantly, the list of sentinels should be seen as a projected list and not a fixed set, evolving over time as new methods are developed and priorities shift.

**Meteorology.** The lower tropospheric structure in the central Arctic is globally distinctive due to the perennial ice pack at its lower boundary. One prevalent and widespread Arctic feature is surface-based temperature inversions, with depths that tend to be greater over sea ice compared to coastal areas and that vary with synoptic activity, cloud cover, and proximity to open water (Vihma, 2014). However, the mechanisms governing this variability are not fully understood and may be evolving with regional declines in sea ice. The competitive effects of the change in the cloud radiative effect compared to the alterations of temperature and humidity profiles with varying ice conditions are unclear (Persson and Vihma, 2017). A potential feedback is a heightened temperature contrast between the atmosphere and ocean, destabilizing the atmospheric boundary layer and promoting convective cloud formation (Vihma et al., 2014). Understanding how Arctic precipitation responds to sea-ice changes and enhanced evaporation at the surface will help to constrain projections of seasonal variability of the Arctic hydrological cycle.

**Clouds.** In a warming Arctic, cloud cover could increase predominantly in the lower troposphere and decrease in the immediate near-surface layers just above the ocean, where prevailing temperature elevations induce atmospheric drying, despite a concurrent increase in moisture flux. The seasonal variability of cloud cover may also be affected because it tends to increase, notably in October, displaying a roughly 1-month delay following the sea-ice minimum in September (Maillard et al., 2021). The increase in cloud cover, especially during autumn, exerts a significant impact on surface downward longwave radiation compared to clear-sky conditions (Kay et al., 2016). In the Arctic, contrary to lower latitudes, the warming effect of clouds due to infrared radiation is much larger than their cooling effect due to attenuation of the solar radiation, except during a very short period in summer (Svensson and Mauritsen, 2020). An enhanced Arctic cloud cover, stemming from diminished sea-ice extent, would therefore lead to more longwave radiation trapped and re-emitted, increasing the temperature at the surface. In turn, this increase could further drive sea-ice retreat and intensify Arctic warming (Persson et al., 2017).

**Trace gases.** The Arctic Ocean is a source of trace gases, notably biological volatile organic compounds (BVOCs), including dimethylsulfide, that can form cloud condensation nuclei and influence clouds (Baggesen et al., 2022). Methane and carbon monoxide also have oceanic sources. The Arctic Ocean is a sink for atmospheric $CO_2$ (Yasunaka et al., 2023), however changes in surface ocean properties of the central basins, including stratification and temperature, can impact $CO_2$ uptake capacity (Cai et al., 2010; Else et al., 2013) and ocean acidity (Zhang et al., 2020) as sea ice retreats. Snow on sea ice is also a source of various trace gases, including nitrogen oxides and halogens containing bromine (Falk et al., 2025, and references therein). Nitrogen oxides in snowpack originate from remission of atmospheric nitrogen deposited onto the surface. Atmospheric nitrogen also originates from remote and local anthropogenic sources (AMAP, 2021) or natural sources (e.g., boreal fires; Ardyna et al., 2022). Halogens play an important role during spring when, for example, so-called bromine explosions deplete tropospheric ozone, to low or near-zero concentrations (Fernandez et al., 2024). Oceanic emissions of iodine compounds also contribute to ozone destruction in the Arctic boundary layer (Benavent et al., 2022), although the spatial extent of these phenomena is still an ongoing research area. Changes in halogens may be influencing long-term trends in ozone depletion events and surface ozone in the Arctic (Law et al., 2023). These cycles are also important for the mercury cycle, as climate-induced changes like sea-ice melting can influence the deposition and release of mercury in the Arctic significantly, with some studies showing increasing mercury levels in certain Arctic biota (Morris et al., 2022).

**Aerosols.** The relative importance of local sources of cloud-active aerosols within the sea-ice pack versus those advected over the sea ice from lower latitudes remains uncertain (Raut et al., 2025). Against a relatively clean background state, transient variability occurs with episodic pulses of aerosols transported into the Arctic and accumulated during winter (Arctic haze). Consequently, the interplay between large-scale meteorology and the persistent near-surface Arctic inversion affects the mixing state, aging, vertical structure, and cloud activity of aerosols, shaping their distribution across the Arctic (Petäjä et al., 2020). The relative contributions of anthropogenic and natural aerosol particles will change over time (Schmale et al., 2022) given the likely increase in biogenic emissions from marine and terrestrial sources (e.g., boreal fires, high-latitude dust). Central Arctic sources of bioaerosols might change due to changes in lead, ridge and melt pond distributions and timing (Ekman and Schmale, 2022; Beck et al., 2024, Schmale et al., 2025). Local anthropogenic sources could also increase with declining sea ice (Raut et al., 2022). Long-term measurements of aerosols and their impacts on pollutant deposition over sea ice or open ocean are lacking, resulting in limited understanding of their variation across the Arctic and their interactions with the climate system and ecosystem functioning, particularly in winter.

**Snow cover.** The snow cover on sea ice influences most surface properties, notably albedo. Accumulation, wind redistribution, and metamorphism are the main processes that determine snow cover (Sturm and Massom, 2017). Snow cover on sea ice is highly heterogeneous spatially, on scales from centimeters to entire floes, and is highly dynamic on time scales of hours to seasons (Webster et al., 2014; 2022). Snow melt causes and controls melt pond formation and impacts the atmosphere-ice-ocean interaction in summer (Niehaus et al., 2023; Smith et al., 2023). In recent decades, snow cover on sea ice has thinned, with melt occurring earlier in spring and accumulation starting later in autumn, although the magnitude and timing of these changes vary strongly across regions. The representation of snow processes in climate models is known to be overly simple, inducing large uncertainties in future snow changes and their consequences on sea-ice mass balance and interactions with light, in particular the transmission of energy to the sea ice and the ocean underneath.

**Ice cover.** Sea ice is composed of different sea-ice types and stages of development over the season (Nicolaus et al., 2022). It is highly heterogeneous on scales of meters to Arctic-wide distribution patterns. Sea-ice dynamic and thermodynamic processes continuously change the volume of sea ice and ice pack properties, altering the fractions of ice types like new ice (refrozen leads), level sea ice, and ridged and deformed sea ice (e.g., Sumata et al., 2023; Von Albedyll et al., 2022). The decrease in area and thickness of the sea-ice cover is the most obvious manifestation for the changing Arctic (Stroeve and Notz, 2018). Changing surface albedo is a key contributor to Arctic amplification and strongly impacts the dispersion and scattering of sunlight (Nicolaus et al., 2012; Pistone et al., 2019). The ice cover is predicted to vanish in summer between 2030 and 2050, leaving an ice-free Arctic for several weeks to months (Notz and SIMIP community, 2020; Jahn et al., 2024). However, many detailed aspects in these predictions are largely uncertain (e.g., ice thickness, light supply to the ocean), resulting from limited process understanding of positive and negative feedbacks and their interplay within the coupled Arctic system.

**Ice dynamics.** Sea ice, a highly dynamic medium, drifts and deforms driven by wind and ocean currents. Pressure ridges and leads form from convergent, divergent and shear movements of pack ice floes. These dynamic processes play an important role for the sea-ice mass balance, in particular from autumn to spring (von Albedyll et al., 2022). Pressure ridges can be much thicker than the surrounding undeformed sea ice, which is important for the dynamic interaction of sea ice with the atmosphere and the ocean. At the same time, pressure ridges are habitats for ice-associated organisms across all trophic levels (Assmy et al., 2013).

Ice velocity has increased over the recent decades (Sumata et al., 2023), influencing sea-ice conditions along the Transpolar Drift (Krumpen et al., 2019). The frequency of ridges is decreasing with younger sea ice (Krumpen et al., 2025), but impacts and future predictions have large uncertainties. Beyond the potential impact on the climate and ecosystem, changing ice dynamics (opening of leads, reduction in ridges) may affect shipping in the Arctic. With TPS, we will contribute to an improved understanding of sea-ice dynamics.

**Ice habitat.** The habitat for organisms living in sea ice and the uppermost ocean is controlled by physical properties and the horizontal and vertical distribution of snow cover, ice cover and underlying water mass properties (e.g., Assmy et al., 2013; Fong et al., 2024; Vancoppenolle et al., n.d.). According to the variability and seasonality of these components, conditions are highly heterogeneous and variable in space and time (Deming and Collins, 2017). As a consequence, the conditions within the ice habitat are not well understood and difficult to quantify and generalize (Campbell et al., 2022). Changes in future ice habitat conditions will be driven by local processes and large-scale processes. For example, the light and energy transfer into and through the snow and sea ice are expected to increase, which will alter the ecological energy budgets (Ardyna and Arrigo, 2020). Further, changes in sea-ice types, thickness, the onset of melt, and formation will affect habitat conditions and consequently ecosystem functioning, structure and biological interactions.

**Ocean hydrography.** Hydrographic and stratigraphic changes in the Arctic Ocean are key indicators of climate change, exhibiting significant spatial heterogeneity over varying timescales. Inflows contribute to the warming of the upper Arctic Ocean and affect freshwater content and stratification (Polyakov et al., 2017; 2020a; Woodgate and Peralta-Ferriz, 2021). Understanding these links requires comprehensive, three-dimensional monitoring of hydrographic properties and sea ice. Enhanced data on regional and seasonal variability are crucial for a pan-Arctic perspective on the physical, geochemical, and biological states of the upper Arctic Ocean. The acquisition of such data is also a prerequisite for the improvement of understanding: i) where and how oceanic heat is being lost, and will be lost in the future, and how such loss will impact sea ice, atmospheric circulation and climate; ii) the current and most likely future state of the halocline (Muilwijk et al., 2023) and its impact on ocean ventilation and vertical fluxes; and iii) the main drivers of the observed changes, including forcing patterns generated by the Arctic Dipole (Polyakov et al., 2024) or the Arctic Oscillation (Morison et al., 2021). Enhanced knowledge on density distribution will improve understanding of the Arctic Ocean's dynamic balance and transfers of ocean properties (Spall, 2020).

**Ocean dynamics.** Arctic Ocean dynamics and eddies play a major role in redistributing physical and biogeochemical properties through ocean currents and mixing, impacting the response of marine ecosystems to climate change. Basin-averaged vorticity balance in the eastern and western Arctic suggests eddies counteract wind-driven vorticity input at the surface, with increasing eddy activity stabilizing the system. However, the timescales for Arctic Basin circulation adjustments remain unclear (Timmermans and Marshall, 2020). Mesoscale eddies transport properties from basin boundaries to the interior, ventilating the halocline (Spall et al., 2008), and may also enhance diapycnal mixing (Rippeth and Fine, 2022), possibly transporting AW heat to sea ice in winter. While eddies are thought to be ubiquitous in the Arctic Ocean, their distribution is largely inferred from models or satellite observations. Intensified eddy activity has been observed in the Beaufort Gyre (Manucharyan et al., 2022), with potential future increases in eddy kinetic energy (Muilwijk et al., 2024). Systematic surveys of eddy features are urgently needed to better understand their role in the redistribution of Arctic Ocean properties, the leading processes underpinning their formation and lifetime, and their response to Arctic warming.

**Ocean mixing.** A pan-Arctic evaluation of mixing in the Arctic Ocean has not been carried out to our knowledge. Frequent mixing, which is linked to strong winds, occurs near topographic features. Lateral thermohaline intrusions near the AW core transport heat to the upper ocean, where staircases (of double diffusion) are observed at the top and bottom interfaces of the intrusions. These features may span larger distances at the upper boundary of the AW layer under low levels of turbulence (Polyakov et al., 2019). Ocean mixing is considered to be the main mechanism for vertical heat transport in the CAO. Near-inertial waves also contribute to stirring properties and are expected to be enhanced under declining sea ice, with reinforced momentum transfer to the ocean and shear-driven turbulent mixing (Polyakov et al., 2020b). Although trends in the level of Arctic Ocean mixing are not yet evident (Lenn et al., 2022), ongoing changes in sea-ice volume, heat and freshwater content, stratification and water masses will affect forcing via mixing and other processes, ultimately having an impact on the water column and sea ice. Key priorities for basin-scale Arctic Ocean mixing observations include developing technologies for sustained, wide-scale ocean microstructure monitoring to better understand the under-ice boundary layer, buoyancy effects of sea-ice growth/melt, momentum transfer through ice, and surface gravity wave generation and propagation.

**Under-ice light.** Arctic warming leads to a reduction in sea-ice thickness and age, as well as an earlier transition from snow-covered ice to ice covered by melt ponds. These factors directly affect surface albedo and the underwater light field. Together, they lead to a substantial increase in the availability of photosynthetically active radiation (400–700 nm) for primary production in the ice-covered upper ocean, promoting under-ice phytoplankton growth and widespread under-ice blooms (Ardyna and Arrigo, 2020). In addition, other evidence indicates that the photosynthetic machinery of Arctic phytoplankton and ice algae can adjust to very low ambient light, enabling positive phytoplankton growth at the end of winter (Kvernvik et al., 2020; Randelhoff et al., 2020; Hoppe et al., 2024) and quick adjustment to the higher light conditions at ice edges (e.g., leads and polynyas; Palmer et al., 2011). In addition, photosensing and photoregulation are likely to play a key role in controlling phytoplankton responses to extreme and variable light conditions (intensity/photoperiod) in the CAO. A better understanding, on both large and small scales, of the changing and variable ice- and light-scape, as well as the unique photosynthetic properties and plasticity of Arctic primary producers, is essential.

**Nutrient availability**. Dissolved nutrients are the fundamental building blocks of marine life, and their availability in the upper Arctic Ocean is generally considered limiting for primary production (Tremblay et al., 2015). While several elements are required to build algal biomass, nitrogen is considered as the main limiting nutrient across most of the Arctic Ocean (Tremblay and Gagnon, 2009), with iron possibly having a role in specific areas (Taylor et al., 2013; Rijkenberg et al., 2018) and silicate limitation being specific to diatoms in the Atlantic sector (Krause et al., 2019). Several knowledge gaps remain with respect to current and future nutrient supplies to the upper water column, which have vertical and horizontal components. These gaps include nutrient exchanges between the brine channels in sea ice and underlying seawater, and microbial processes in the ice and water column that modulate the bioavailability of nitrogen. The magnitude and composition of horizontal nutrient inputs via the Bering Sea, the Barents Sea Opening, rivers and coastal erosion are strikingly different and susceptible to an array of remote external forcings that are poorly constrained (e.g., Tremblay et al., 2015; Terhaar et al., 2021). Convection and the vertical export of organic matter at shelf edges transfer these external nutrients into the halocline where they can remain isolated from the surface ocean. Their upward re-supply in the Arctic Ocean's interior is contingent on vertical mixing and entrainment processes that remain sparsely measured and coarsely modelled (e.g., Randelhoff et al., 2020; Rippeth and Fine, 2022). In parallel, microbial processes occurring both in the water column and in sea ice, affect the inventory of bio-labile nitrogen through net gains (e.g., $N_2$ fixation) or losses (e.g., denitrification, anammox) that remain poorly quantified (Vancoppenolle et al., 2013; Von Friesen and Riemann, 2020; Zhuang et al., 2022).

**Ocean biodiversity.** The CAO is home to endemic fauna that have evolved with a multi-year ice cover, cold temperatures, an extreme light regime, and pulsed productivity. Changes in these extreme conditions result in the borealization of the Arctic fauna, defined as a process by which ecosystems that were characterized historically as Arctic are progressively acquiring features typical of more southern, boreal ecosystems (Oziel et al., 2019). Knowledge gaps persist with regards to the rate of borealization in the CAO and resulting impacts on structure and functionality of food webs. Whether diversity would increase or decrease due to niche partitioning or disappearance of endemic species and whether borealization happens gradually or by drastic changes in ecosystem dynamics are both unclear. Investigating these aspects is critical to assess the productivity and biodiversity of the future CAO ecosystems. We predict that, although total biodiversity measured by species numbers might increase, genomic biodiversity could decrease if sympagic fauna decrease across all seasons, including winter (Ehrlich et al., 2020). Adaptation to new conditions and competition with boreal taxa may also result in altered life histories of Arctic species, including timing of reproduction, grazing, and vertical distributions (Moline et al., 2008). Moreover, Hop et al. (2020) reported that reduction in multi-year ice could result in lower protist biodiversity in the CAO.

**Microbial functioning.** Microbial ecosystem functioning in the Central Arctic Ocean (CAO) arises from the combined activity of microorganisms inhabiting the sea ice, the underlying water column, and the ocean interior. These assemblages include photosynthetic, heterotrophic, and chemoautotrophic taxa that together sustain the marine food web and drive key biogeochemical processes regulating carbon and nitrogen cycles (Falkowski et al., 2008). Ongoing warming and sea-ice loss in the CAO are expected to affect the composition and metabolic activity of both pelagic and sympagic microorganisms, thereby altering microbial functional diversity and the overall ecosystem processes they sustain. Heterotrophic or chemoautotrophic bacteria or archaea capable of fixing carbon could be examples of microorganisms in the CAO that continue to fix biological carbon during the long months of darkness, when the activity of photosynthetic algae is highly reduced or absent (Alonso-Sáez et al., 2010). Specific microbial functions can be identified using diagnostic genes encoding targeted metabolic pathways. For example, detecting transcripts of genes involved in $CO_2$-fixation metabolism in prokaryotes could indicate which microbes perform this function and their persistence in the CAO. Following this approach, suites of marker genes representing essential ecosystem processes could be systematically identified and monitored over time to track microbial functional dynamics in the rapidly changing CAO.

**Carbon export.** Increased advection of Atlantic and Pacific waters into the deep basins of the CAO is altering sea-ice characteristics, nutrient availability, and the timing and intensity of the productive season (Wassmann and Reigstad, 2011; Lewis et al., 2020). These shifts affect not only the magnitude of primary production but also its phenology and spatial distribution (Ardyna and Arrigo, 2020). Changes in sea-ice cover and snow properties further influence light availability, stratification, and the vertical transfer of nutrients, thereby modifying key biogeochemical cycles and carbon fluxes (Nöthig et al., 2020; Ramondenc et al., 2023). How ongoing transformations in both sympagic and planktonic communities, and the changing coupling between them, particularly shifts in diversity, size spectra and abundance, will affect the quantity, composition and sinking velocity of exported material in the CAO remains unclear. Similarly, the long-term implications of changing lipid production and transfer through evolving food webs for carbon flux dynamics remain unknown. Whether the CAO will exhibit export fluxes more comparable to productive shelf regions in the seasonal ice zone, or whether increased stratification and oligotrophication in the deep basins will suppress vertical carbon export, is a critical open question. The fate of carbon in this rapidly changing region has important implications for global carbon budgets and life on the seafloor (e.g., Boetius et al., 2013). To determine whether the deep basins of the CAO will evolve into a significant and persistent carbon sink on climate-relevant timescales, or if biological export pump contributions will remain marginal under future Arctic conditions, is thus essential.

**Mercury.** Elevated concentrations of mercury (Hg) are found in Arctic biota, potentially putting both ecosystem and human health at risk (Basu et al., 2022; Dietz et al., 2022). The Arctic

Ocean is unique in exhibiting elevated surface Hg and shallow methylmercury peaks (Heimbürger et al., 2015). Hg enters the system via atmospheric deposition, riverine inflow, coastal erosion, and oceanic exchange (Dastoor et al., 2022). Major sinks include particle-driven vertical export on shelves (Tesan et al., 2020) and outflow to the Atlantic (Petrova et al., 2020). The atmosphere plays a central role, with springtime atmospheric mercury depletion events causing deposition on snow and ice (Ahmed et al., 2023), followed by substantial summer remission, making net deposition highly variable. Evasion from open leads and the marginal ice zone is also a major atmospheric source (Yue et al., 2023). Seasonal and interannual variability remains poorly constrained due to limited year-round observations (Angot et al., 2022; Kohler et al., 2024). The short residence time of Hg (<10 years) makes the Arctic Ocean particularly sensitive to change (Kohler et al., 2022). Despite global emission reductions, thawing permafrost is expected to release legacy Hg (Chételat et al., 2022), while warming likely enhances microbial methylation and bioaccumulation (McKinney et al., 2022). The transition from multi-year to first-year sea ice may further elevate methylmercury concentrations within the sea-ice habitat (Schartup et al., 2020).

**Plastics.** A recent review of over 60 studies on plastic pollution in the Arctic found that macroplastic objects (>5 mm) and microplastic fragments (<5 mm) are pervasive throughout the Arctic, even in areas with no apparent human activity, such as the CAO and the deep seafloor (Bergmann et al., 2022). Some plastic pollution is from local sources, such as fisheries, landfills, wastewater and industrial activity, but long-range sources are also substantial, as plastic is transported from the mid-latitudes to the Arctic by Pacific and Atlantic waters, by atmospheric transport and by the large watersheds of Arctic rivers (Materić et al., 2022). As global plastic production is set to double by 2050 (Lebreton and Andrady, 2019), we expect increased transport of microplastics to Arctic sea ice and the larger marine ecosystem. Small, floating microplastics enter the food web by assimilation, ingestion or inhalation, and they release toxic additives and monomers inside organisms and in the environment (Hermabessiere et al., 2017). The consequences for Arctic wildlife are a greater health burden, including disrupted endocrine function and increased risk for neurodevelopmental disorders, reproductive birth defects, and infertility (Gallo et al., 2018).

**Other chemicals of emerging Arctic concern**. In addition to Hg and plastics, a growing number of persistent and mobile synthetic chemicals are raising concern in the Arctic due to their long-range transport potential, environmental persistence, and bioaccumulation in food webs (AMAP, 2017). For example, perfluoroalkyl and polyfluoroalkyl substances (PFAS), widely used in consumer products, industrial processes, and firefighting foams, are now detected frequently in Arctic air, snow, surface waters, and biota, including top predators (Xie et al., 2022). Their presence in remote Arctic environments highlights the efficiency of atmospheric and oceanic transport pathways. These emerging organic contaminants have been associated with immunotoxicity, endocrine disruption, and developmental toxicity in both wildlife and humans (Sonne et al., 2021). Given ongoing chemical innovation, the number of contaminants reaching the Arctic is likely to increase, posing new challenges for monitoring and risk assessment in a rapidly changing environment.

## 4. Approach

The guiding principle for the Tara Polaris observational program is to measure basic parameters for each of the three compartments (i.e., atmospheric, sea ice/snow and ocean) over the course of two decades. Table 2 lists these basic parameters to be measured and observed consistently on all drifts; Figure 3 illustrates the basic measurement approaches, with TPS in the center of activities. These measurements constitute the backbone of the observation concept, with spatial and temporal resolution limited by the restricted resources on board. The timing and spatial coverage of observations could be adjusted over time, as one of the primary goals is to identify the relevant scales of variability (from sub-daily to inter-annual) in the context of global change. Methods are chosen to be as autonomous as possible. Manual sampling and processing are included where necessary, particularly for ecological parameters. In addition to these compartment-specific protocols, several transversal approaches, notably contaminant analyses and microbial omics, will be applied systematically across the atmosphere, sea ice/snow, and ocean compartments. These integrated measurements will ensure methodological comparability, facilitate data synthesis across compartments, and strengthen the overall capacity of TPS to capture coupled physical and biogeochemical processes.

At the same time, Figure 4 shows that the observational program will be realized such that measurements overlap as much as possible to allow best linkages of observations across the compartments and related parameters. Additional parameters and higher-frequency measurements are implemented as part of targeted studies, as described separately for the atmospheric (Schmale et al., 2025), sea ice and snow (Vancoppenolle et al., n.d.), and ocean (Geoffroy et al., n.d.) components.

Each drift will be unique in terms of its trajectory and weather conditions. The main approach is to carry out repeated drift experiments along the transpolar drift, following the ice cover from its early formation into melt. However, targeted drift scenarios can be chosen for individual years, focusing on specific processes. This dual strategy makes it possible to quantify the relative importance of processes and links between compartments (i.e., atmosphere, sea ice/snow and ocean). Observation of processes, such as fluxes across interfaces and linkages between subsystems, will improve their parameterization in models and lead to a better understanding of the Arctic system and its evolution beyond 2046.

## 4.1 Atmospheric measurements

Long-term basic meteorological observations, including pressure, temperature, humidity, wind speed, and wind direction, will be gathered by two atmospheric weather stations: one onboard TPS and another on a mast situated over the sea-ice pack. These observations will be complemented by regular measurements of precipitation intensity and $CO_2$ concentration in air. While eddy-covariance measurements are currently the state of the art for determining air-sea $CO_2$ fluxes, use of such a system in harsh polar environments is especially challenging (e.g., Butterworth and Miller, 2016), therefore we will continue to develop this capacity for future TPS campaigns. In the meantime, continuous observations of $CO_2$ concentration in air and surface waters (see 4.3), combined with meteorological and oceanographic parameters (4.3), will enable us to *estimate* air-sea $CO_2$ fluxes and provide essential context to develop future dedicated $CO_2$ flux installations. Additionally, shortwave and longwave upwelling and downwelling radiation (from broad-band radiometers), as well as sensible and latent turbulent fluxes, will be measured to constrain the surface energy budget, in conjunction with conductive heat fluxes reported in the ice.

To gather data on airborne microorganisms, aerosol chemical composition and their sources particles will be collected regularly on filters both on TPS and at remote locations over sea ice. General aerosol information, such as their mass, concentration, number, and size distributions, as well as trace gas (ozone, carbon monoxide, BVOCs like dimethyl sulfide using cartridges) and contaminant (Hg) concentrations, will be monitored continuously using instruments permanently installed on TPS. Measurements of gaseous nitrogen compounds such as nitric oxides or nitric acid and halogens will require dedicated measurement efforts. In addition to aerosol microphysical and optical properties (scattering/absorption), basic aerosol sampling on filters will permit the study of chemical composition, including black carbon, organics, soluble inorganics, methanesulfonic acid, microplastics and mineral dust. Their potential activity to serve as ice-nucleating particles will also be investigated, together with measurements of cloud condensation nuclei.

To address the vertical dimension, we will first deploy remote sensing instruments (lidar, radar and zenith-sky UV-visible spectrometer) onboard TPS to detect aerosol optical depth, cloud structures, optical depth and phase partitioning, and UV radiation fluxes. Enabling horizontal scanning observations of the lidar system will permit the documenting of spatial variability of aerosols over the heterogeneous structures of sea ice, including leads, melt ponds or ridges. In addition, a small tethered balloon and drones will be used to obtain vertical and horizontal in-situ data, respectively. For more detailed investigations (see also Schmale et al., 2025) into the processes governing surface fluxes (emission, deposition) and microorganism-cloud interactions, we plan to conduct dedicated field campaigns with more comprehensive measurements. These efforts may include, for example, deploying a large tethered balloon or taking measurements from a sea-ice-based mast to study deposition fluxes.

### 4.2 Sea-ice and snow measurements

Parameters for sea ice and snow mass (including sea-ice and snow thicknesses, freeboard, sea-ice and snow densities) and energy balance (including sea-ice and snow temperatures, surface albedo, light transmittance, and surface conditions) will be measured with autonomous platforms fixed to buoys on different ice types and manually along transects in the vicinity of TPS. Sea-ice draft and bottom roughness as well as under-ice optical and water mass properties will be mapped in two dimensions (e.g., 200 m radius around TPS) with different sensors mounted on a remotely operated vehicle on a weekly basis. Surface conditions, including melt pond coverage in summer, will be observed visually using cameras installed on autonomous stations, on TPS, and on drones.

Physical and ecological properties of sea ice and snow, including salinity, density, stratigraphy, texture, and chlorophyll *a* fluorescence, will be measured from snow pit and ice core samples on one or two sites on the ice. The composition of ice algae and other ice-inhabiting organisms, as well as phytoplankton and zooplankton communities, will be sampled at the ice-ocean interface, providing data on the biodiversity and spatial heterogeneity of the sea-ice habitat. This sampling will be aligned with the measurements of impurities (e.g., black carbon, dust, and other light-absorbing particles), nutrients, particulate matter and contaminants (Hg, microplastics, and other chemicals of emerging concern in the Arctic). Under-ice and ice bottom samples will be obtained using the remotely operated vehicle. Additional samples may be collected for incubations and experiments related to analyses in the atmosphere, sea-ice/snow and ocean compartments.

### 4.3 Ocean measurements

To obtain the basic information about hydrography, regular conductivity-temperature-depth (CTD) casts, including standard additional instrumentation (dissolved oxygen, photosynthetically active radiation, turbidity, biogeochemical sensors) will collect profiles from the surface to 2000 m, thus covering the water column to the approximate depth of the Lomonosov Ridge (which separates the Amerasian Basin from the Eurasian Basin). Profiles to the bottom will be carried out over longer time intervals to monitor long-term changes in the deep Arctic Ocean water masses. Current velocity profiles will be monitored in the upper 800 m from a station-mounted Acoustic Doppler Current Profiler (ADCP) and larger plankton will be collected using nets deployed from the moon pool. For the upper water column, under the TPS, discrete water samples will be collected using the rosette, while surface water from 2.5 m depth will be pumped through a seawater intake for continuous monitoring of temperature, salinity, dissolved oxygen, $pCO_2$, pH, chlorophyll *a* fluorescence, and bio-optical properties. Complementary measurements to characterize the unperturbed top ocean layer away from the station will be maintained from bore holes in the ice near the TPS by deploying sediment traps, ADCPs with turbulence-measurement capacity, a light chain to measure irradiance, and an echosounder for biological backscatter and sea-ice bottom profiling.

Sampling under the ice beyond the TPS footprint will include CTD chains with ADCP and turbulence sensors, short- and long-term sediment trap deployments with underwater vision profilers, an autonomous high-frequency echosounder (Acoustic Zooplankton and Fish Profiler), and ichthyoplankton net deployments. Sample collection will combine direct measurements and laboratory analyses, including state-of-the-art omics, culture and imaging approaches (e.g., long-read sequencing for metagenomics and single-cell genomics), as well as the quantification of selected contaminants.

**4.4 Context to other approaches**

All measurements and observations will be limited to a radius of a few kilometers around TPS along the specific drift trajectory. Collaboration and coordination with other programs will be necessary to contextualize all the collected data, assure compatibility of protocols for comparisons, and extend spatial scales and temporal resolution. While extensive collaboration is planned for all drifts, a special effort is expected for the International Polar Year (2030–2032).

**Other vessels.** A number of nations carry out scientific expeditions with research vessels in the CAO. These expeditions often provide interdisciplinary atmosphere-ice-ocean measurements. Unlike the continuous drift of TPS in the same sea ice throughout the year, these expeditions traverse mainly the ice-covered Arctic Ocean. As a result, these measurements often add a spatial component within the same year. The TPS consortium will be in close contact with these expeditions to share observational concepts, exchange positions and activity information, discuss results and coordinate measurements wherever possible.

**Buoy networks and moorings.** The autonomous stations (buoys) at TPS will be shared with the International Arctic Buoy Program and data made available in near real time, e.g., through this buoy program and the online sea-ice knowledge and data platform (meereisportal.de; Grosfeld et al., 2016). Complementarily, the data from all the other buoys will provide the same type of data from other regions in the Arctic. Similarly, TPS will drift in regions where moorings are installed, providing stationary time series in addition to the drifting time series. By merging all these datasets, a large Lagrangian dataset (from the sea-ice perspective) describing atmosphere-snow-ice-ocean conditions will become available.

**Ground-based observatories.** Continuous, long-term atmospheric measurements are collected as part of the International Arctic Systems for Observing the Atmosphere network, which provides info on the Arctic atmospheric environment from its stations across the Arctic (Uttal et al., 2016). Observatories are more concentrated in the western Arctic than in the eastern Arctic: Utqiagvik (Alaska), Eureka and Alert (Canada), Ny-Alesund, Zeppelin and Andoya (Norway), Summit Station, Zackenberg/Daneborg and Villum (Greenland), Pallas and Sodankylä (Finland), and Tiksi and Cherski (Russia). This network is designed to provide a comprehensive understanding of Arctic atmospheric processes and their role in the global climate system. Typically, these network stations collect data on meteorology, radiation, greenhouse and other trace gases, clouds and precipitation, and aerosol particles, enabling assessment of aerosol trends and policy impact as part of the Arctic Monitoring and Assessment Program (AMAP, 2021). Although collectively crucial for observing long-term trends and changes in Arctic climate, these stations are more representative of the Arctic coastal environment than its central ocean. The Tara Polaris expeditions provide a unique opportunity to fill the gaps in atmospheric observations in the CAO.

**Aircraft.** TPS is expected to be visited occasionally by aircraft and helicopters from other vessels. Their longer-range measurements of the state of the atmosphere, snow and sea ice will be complementary to the local measurements taken at TPS in a spatial context. They will contribute to discussions on the representativeness and spatial variability of several long-term sentinels.

**Satellites.** Surface measurements and TPS observations offer an excellent opportunity to establish a link with satellite observations. Ground measurements will be used to improve satellite retrievals, while satellite data will place TPS observations on a larger spatial scale. The set of parameters derived from satellite measurements is extensive, covering most of the long-term sentinels of atmospheric and sea ice measurements (Table 2). However, limitations will apply to the periods during which TPS will drift into the polar hole of different satellites, and during polar night for passive optical sensors.

**Numerical modeling.** The most interaction between Tara Polaris measurements and numerical modeling is anticipated for process-oriented models in the three compartments. The Tara Polaris data will be used for model initialization, forcing and validation to improve the understanding of processes and support further models towards better forecasts and predictions of changes in climate, biogeochemical cycling and ecosystems. Regional models, in particular weather and ice forecasts, will also support TPS operations.

All these collaborative approaches will support the planning and execution of each drift. The information they provide will be invaluable to anticipating conditions and thus measurements, scales and resolutions to prioritize or consider adjusting.

**4.5 From observation to archive**

One of the main challenges of this 20-year observation program is to guarantee consistency and excellence throughout, as most sampling and measurements will be carried out by a small and changing group of people on board. Tara Polaris will establish new procedures for the entire chain, from observations to archives, to: 1) define measurement and metadata protocols for each parameter and method; 2) pre-train the responsible team on board; 3) support the team during the drift; 4) carry out measurements and sampling during the drift; 5) document measurements and communicate data samples for quality checks, automatically or manually, to the home institutes; 6) store all data in a structured project database; 7) calibrate, quality-control and apply standard data processing to all data and merge with metadata; and 8) publish and archive data according to the FAIR principles of Findable, Accessible, Interoperable, and Reusable. This process and data storage must follow existing national and international infrastructures. The decision to transmit data in real or delayed time must be taken for each dataset.

The Tara Polaris dataset will need to reach a new level of data processing due to the extreme heterogeneity of the data at different spatial scales (from point measurements and individual samples, to time series and transects, up to satellite grids), different temporal resolutions (from seconds to daily and monthly datasets) and different data types (numerical data, omics data, images). The aim is to obtain weekly data sets that can be used to generalize atmosphere-ocean-ice conditions along the drift trajectory.

# 5. Foreseen outcomes

TPS will deliver the first multi-decadal record of the coupled atmosphere–ice–ocean system in the central Arctic, enabling a leap in our ability to separate natural variability from long-term climate change. Previous large-scale efforts, such as SHEBA, DAMOCLES, MOSAiC, and others cited in the introduction, have provided invaluable insights into seasonal and interannual processes. Yet, by their very design as 1-year or short-term expeditions, they capture only a "snapshot" of an extremely dynamic system. The TPS will extend these insights by establishing a sustained 20-year baseline, which is essential to i) document processes with variabilities that unfold over the decadal scale, ii) capture rare or extreme events that occur only sporadically; and iii) provide an integrated reference dataset against which models and satellite missions can be tested systematically and improved. In short, TPS will transform the way we observe and understand the Arctic by combining long-term continuity with system-level integration.

**Atmosphere.** The Arctic atmosphere exerts a fundamental control on the regional energy budget, cloud radiative forcing, and chemical composition, yet it remains one of the least well-constrained components of models of the climate system. TPS will generate for the first time a continuous, multi-decadal record of Arctic atmosphere-ice interactions. This record will make possible the quantification of decadal variability in stratification of the lower atmosphere, radiation and cloud processes, such as the frequency and intensity of moist intrusions, fog, and atmospheric rivers that episodically dominate the polar energy budget. It will also provide a tracking of long-term changes in aerosols, air-sea $CO_2$ fluxes, trace gases and pollutants, including black carbon, tropospheric ozone, methane, and other short-lived climate forcers, which have origins and radiative and biogeochemical impacts, including deposition, that are only poorly represented in models. Moreover, TPS will assess the interplay between biology and aerosols in the atmosphere, by linking local emissions from leads, melt ponds, and microbial activity with regional cloud formation and atmospheric composition. Finally, the extended record will capture extreme and rare events, such as winter warming spikes, ozone depletion events, or anomalous transport episodes, which can be fully contextualized only with a multi-decadal dataset.

**Snow and sea ice.** Sea ice and its snow cover are at the heart of Arctic climate feedback, controlling albedo, heat fluxes, and ecosystem functioning, yet both remain poorly parameterized in models and imperfectly constrained by satellite observations, largely because of their strong spatial heterogeneity and rapid evolution. TPS will provide the first long-term record of snow and sea-ice properties from a drifting platform, enabling us to document the evolution of snow depth, distribution, and metamorphism and how they shape surface albedo and the energy balance of the Arctic system. TPS will also track the transition from multi-year to first-year ice, including associated changes in ice thickness and other characteristics, including roughness, melt pond dynamics, and permeability. With this continuity, TPS will assess the balance between short-term variability and structural regime shifts in the ice cover, disentangling transient extremes from long-term trends. Finally, this dataset will provide the empirical basis to improve the representation of snow and ice in climate and pollutant cycling models and to validate and refine satellite retrievals of sea-ice thickness and snow depth.

**Ocean.** The Arctic Ocean is undergoing rapid and profound transformations in stratification, circulation, and ecosystem functioning, yet decadal-scale changes remain essentially unknown. TPS will establish the first multi-decadal baseline of the upper Arctic Ocean, making possible the quantification of long-term changes in mixed-layer depth, freshwater storage, and stratification, and their role in modulating nutrient and pollutant fluxes and upper-ocean dynamics. This record will also track biogeochemical variability, including nutrients, pollutants, dissolved gases, carbonate chemistry, and air–sea $CO_2$ fluxes, with a focus on emerging signals of acidification and deoxygenation. At the same time, TPS will assess the response of communities, primary production, and trophic interactions to shifting sea-ice cover and stratification, with particular attention to under-ice blooms and export fluxes. Ultimately this continuity will provide a robust multi-decadal baseline to constrain coupled physical–biogeochemical models and evaluate ecosystem resilience under ongoing Arctic change.

**Integration and legacy.** What sets TPS apart is not only the continuity of the record, but also the simultaneous observation of all components of the Arctic system. This integration will allow researchers to identify cross-system linkages, capture cascading effects, and quantify the resilience and thresholds of the Arctic climate system. By spanning two decades, TPS will cover multiple phases of major climate oscillations (North Atlantic Oscillation, Atlantic Multidecadal Oscillation, El Niño Southern Oscillation) and capture their integrated impacts on the Arctic system and potentially vice versa. This holistic approach will provide an unparalleled testbed for models and satellite algorithms, bridging the gap between short-term field campaigns and long-term monitoring networks. Beyond the scientific advances, TPS will establish a legacy dataset that will serve as a benchmark for future generations to assess Arctic change, ensuring that the Arctic, as the most rapidly changing component of the Earth system, is also the best observed — a global bellwether for climate change.

# 6. Concluding remarks

The Tara Polaris multi-decadal observation program represents an unprecedented opportunity to document, understand, and anticipate the profound transformations occurring in the CAO. By combining continuous, year-round measurements across atmosphere, sea ice and snow, and ocean compartments with integrative analyses and cutting-edge methodologies, the program will generate a unique legacy dataset. This long-term effort is essential to disentangle natural variability from anthropogenic trends, assess feedback within the coupled Arctic system, and constrain model projections in a region undergoing rapid and nonlinear changes.

Beyond scientific advancement, Tara Polaris has the potential to redefine how sustained Arctic observations are conducted. Its modular, low-impact platform offers a flexible, replicable model for future polar monitoring initiatives. Through open science, interoperability, and partnerships with international stakeholders (e.g. European consortium), Tara Polaris will reinforce the Arctic's role as a sentinel of global change. Ultimately, this program aims to strengthen scientific diplomacy, inform policy frameworks, and support a collective response to the climate and ecological crises unfolding in the Arctic and beyond. It will also serve as a powerful vehicle for science outreach and public engagement, raising awareness about the Arctic's transformation and fostering broader societal mobilization through education, storytelling, and participatory initiatives.

# 7. Acknowledgment

M.A. received funding from the ANR Arclight (grant ANR-25-CE20-4776), Horizon Europe programme under Grant Agreement No 101136875 via project POMP (Polar Ocean Mitigation Potential); from Sentinelle Nord; from ArcticNet, the Natural Sciences and Engineering Research Council of Canada through the Discovery program and CNES Alg-O-Nord.

M.N. was partly funded through the EU Horizon 2020 project Arctic Passion (Grant 101003472), the Deutsche Forschungsgemeinschaft (DFG, German Research Foundation) through the Transregional Collaborative Research Centre TRR-172 "ArctiC Amplification: Climate Relevant Atmospheric and SurfaCe Processes, and Feedback Mechanisms (AC)3" (grant 268020496, project C01), and the German Ministry of Education and Research (BMBF) through the project MOSAiC3-IceScan (BMBF 03F0916A).

J.-C.R. received funding from Horizon Europe programme under Grant Agreement No 101137680 via project CERTAINTY (Cloud-aERosol inTeractions & their impActs IN The earth sYstem); from the European Union's Horizon 2020 research and innovation programme under Grant agreement No 101003826 via project CRiceS (Climate Relevant interactions and feedbacks: the key role of sea ice and Snow in the polar and global climate system) ; and from the French National Research Agency (ANR) via the project MPC2 (n° ANR-22-CEA01-0009-02).

H.A. acknowledges financial support from the ANR ATOX project (grant ANR-24-CE01-7616).

K.B. is funded through the Canada Research Chairs Program.

J.M.F is supported by a research grant from Scott Eric Jordan.

M.G. is supported by the Natural Sciences and Engineering Research Council of Canada through the Discovery program and the SEDNA project funded by the Norwegian Research Council (Grant # 352539).

K.S.L. acknowledges funding from ANR CASPA (Climate-relevant Aerosol Sources and Processes in the Arctic) project (Grant ANR-21-CE01-0017) and CNES MERLIN.

S.R. acknowledges funding from Danish National Research Foundation (grant no. DNRF 185).

J.S. holds the Ingvar Kamprad Chair for Extreme Environments Research sponsored by Ferring Pharmaceuticals. J.Schmale acknowledges financial support via Swiss National Science Foundation project grant no. 212101.

J.E.S. acknowledges the ANR-20-CE34-0014 ATMO-PLASTIC and ANR-23-CE34-0012 BUBBLPLAST grants.

The Tara Ocean Foundation is globally funded on this program by France 2030 in the frame of the French Polar Strategy, BNP Paribas, Cap Gemini Engineering, Fondation Prince Albert II de Monaco, Monaco Explorations, Fondation Albedo pour la Cryosphère, Fonds de dotation O, Fondation Didier et Martine Primat, agnès b., Bureau Veritas and Fondation Veolia.

## 8. Data accessibility

No data have been used in this manuscript.

## 9. References

Ahmed, S, Thomas, JL, Angot, H, Dommergue, A, Archer, SD, Bariteau, L, Beck, I, Benavent, N, Blechschmidt, A-M, Blomquist, B, Boyer, M, Christensen, JH, Dahlke, S, Dastoor, A, Helmig, D, Howard, D, Jacobi, H-W, Jokinen, T, Lapère, R, Laurila, T, Quéléver, LLJ, Richter, A, Ryjkov, A, Mahajan, AS, Marelle, L, Pfaffhuber, KA, Posman, K, Rinke, A, Saiz-Lopez, A, Schmale, J, Skov, H, Steffen, A, Stupple, G, Stutz, J, Travnikov, O, Zilker, B. 2023. Modelling the coupled mercury-halogen-ozone cycle in the central Arctic during spring. *Elementa: Science of the Anthropocene* 11(1): 00129. doi: 10.1525/elementa.2022.00129

AMAP. 2017. Chemicals of Emerging Arctic Concern. Summary for Policy-makers. Arctic Monitoring and Assessment Programme (AMAP), Oslo, Norway. 16 pp

AMAP. 2021. 2021 AMAP Mercury Assessment. Summary for Policy-makers. Arctic Monitoring and Assessment Programme (AMAP), Tromsø, Norway.

Alonso-Sáez, L, Galand, PE, Casamayor, EO, Pedrós-Alió, C, Bertilsson, S. 2010. High bicarbonate assimilation in the dark by Arctic bacteria. *ISME Journal* 4(12): 1581–1590. doi: 10.1038/ismej.2010.69

Anderson, DM, Fachon, E, Hubbard, K, Lefebvre, KA, Lin, P, Pickart, R, Richlen, M, Sheffield, G, Van Hemert, C. 2022. Harmful algal blooms in the Alaskan Arctic: An emerging threat as oceans warm. *Oceanography* 35(2). doi:10.5670/oceanog.2022.121.

Angot, H, Blomquist, B, Howard, D, Archer, S, Bariteau, L, Beck, I, Boyer, M, Crotwell, M, Helmig, D, Hueber, J, Jacobi, H-W, Jokinen, T, Kulmala, M, Lan, X, Laurila, T, Madronich,

M, Neff, D, Petäjä, T, Posman, K, Quéléver, L, Shupe, MD, Vimont, I, Schmale, J. 2022. Year-round trace gas measurements in the central Arctic during the MOSAiC expedition. *Scientific Data* 9(1): 723. doi: 10.1038/s41597-022-01769-6

Ansmann, A, Ohneiser, K, Engelmann, R, Radenz, M, Griesche, H, Hofer, J, Althausen, D, Creamean, JM, Boyer, MC, Knopf, DA, Dahlke, S, Maturilli, M, Gebauer, H, Bühl, J, Jimenez, C, Seigert, P, Wandinger, U. 2023. Annual cycle of aerosol properties over the central Arctic during MOSAiC 2019–2020 – light-extinction, CCN, and INP levels from the boundary layer to the tropopause. *Atmospheric Chemistry and Physics* 23(19): 12821–12849. doi: 10.5194/acp-23-12821-2023

Ardyna, M, Arrigo, KR. 2020. Phytoplankton dynamics in a changing Arctic Ocean. *Nature Climate Change* 10(10): 892–903. doi:10.1038/s41558-020-0905-y.

Ardyna, M, Hamilton, DS, Harmel, T, Lacour, L, Bernstein, DN, Laliberté, J, Horvat, C, Laxenaire, R, Mills, MM, van Dijken, G, Polyakov, I, Claustre, H, Mahowald, N, Arrigo, KR. 2022. Wildfire aerosol deposition likely amplified a summertime Arctic phytoplankton bloom. *Communications Earth Environment* 3(1): 201. doi: 10.1038/s43247-022-00511-9

Arrigo, KR, van Dijken, GL, Cameron, MA, van der Grient, J, Wedding, LM, Hazen, L, Leape, J, Leonard, G, Merkl, A, Micheli, F, Mills, MM, Monismith, S, Ouellette, NT, Zivian, A, Levi, M, Bailey, RM. 2020. Synergistic interactions among growing stressors increase risk to an Arctic ecosystem. *Nature Communications* 11:6255. doi:10.1038/s41467-020-19899-z.

Årthun, M, Eldevik, T, Smedsrud, LH. 2019. The role of Atlantic heat transport in future Arctic winter sea ice loss. *Journal of Climate* 32(11):3327–3341. doi:10.1175/jcli-d-18-0750.1.

Assmy, P, Ehn, JK, Fernandez-Mendez, M, Hop, H, Katlein, C, Sundfjord, A, Bluhm, K, Daase, M, Engel, A, Fransson, A, Granskog, MA, Hudson, SR, Kristiansen, S, Nicolaus, M, Peeken, I, Renner, AHH, Spreen, G, Tatarek, A, Wiktor, J. 2013. Floating ice-algal aggregates below melting Arctic sea ice. *PLoS One* 8(10):e76599. doi:10.1371/journal.pone.0076599.

Baggesen, NS, Davie-Martin, CL, Seco, R, Holst, T, Rinnan, R. 2022. Bidirectional exchange of biogenic volatile organic compounds in subarctic heath mesocosms during autumn climate scenarios. *Journal of Geophysical Research: Biogeosciences* 127(6):e2021JG006688. doi:10.1029/2021JG006688.

Basu, N, Abass, K, Dietz, R, Krümmel, E, Rautio, A, Weihe, P. 2022. The impact of mercury contamination on human health in the Arctic: A state of the science review. *Science of The Total Environment* 831: 154793. doi:10.1016/j.scitotenv.2022.154793.

Beck, I, Moallemi, A, Heutte, B, Pernov, JB, Bergner, N, Rolo, M, Quéléver, LLJ, Laurila, T, Boyer, M, Jokinen, T, Angot, H, Hoppe, CJM, Müller, O, Creamean, J, Frey, MM, Freitas, G, Zinke, J, Salter, M, Zieger, P, Mirrielees, JA, Kempf, HE, Ault, AP, Pratt, KA, Gysel-Beer, M, Henning, S, Tatzelt, C, Schmale, J. 2024. Characteristics and sources of fluorescent aerosols in the central Arctic Ocean. *Elementa : Science of Anthropocene* 12(1):00125. doi:10.1525/elementa.2023.00125.

Benavent, N, Mahajan, AS, Li Q, Cuevas, CA, Schmale, J, Angot, H, Jokinen, T, Quéléver, LLJ, Blechschmidt, A-M, Zilker, B, Richter, A, Serna, JA, Garcia-Nieto, D, Fernandez, RP, Skov, H, Dumitrascu, A, Pereira, PS, Abrahamsson, K, Bucci, S, Duetsch, M, Stohl, A, Beck, I, Laurila, T, Blomquist, B, Howard, D, Archer, SD, Bariteau, L, Helmig, D, Hueber, J, Jacobi, H-W, Posman, K, Dada, L, Daellenbach, KR, Saiz-Lopez, A. 2022. Substantial contribution of iodine to Arctic ozone destruction. *Nature Geoscience* 15(10):808–814. doi:10.1038/s41561-022-01023-z.

Bergmann, M, Collard, F, Fabres, J, Gabrielsen, GW, Provencher, JF, Rochman, CM, van Sebille, E, Tekman, MB. 2022. Plastic pollution in the Arctic. *Nature Reviews Earth Environment* 3(5):323–337. doi:10.1038/s43017-022-00279-8.

Beszczynska-Möller, A, Fahrbach, E, Schauer, U, Hansen, E. 2012. Variability in Atlantic water temperature and transport at the entrance to the Arctic Ocean, 1997–2010. *ICES Journal of Marine Science* 69(5):852–863. doi:10.1093/icesjms/fss056.

Boetius A, Albrecht S, Bakker K, Bienhold C, Felden J, Fernández-Méndez M, Hendricks S, Katlein C, Lalande C, Krumpen T, Nicolaus M, Peeken I, Rabe B, Rogacheva A,

Rybakova E, Somavilla R, Wenzhöfer F, RV Polarstern ARK27-3-Shipboard Science Party 2013. Export of Algal Biomass from the Melting Arctic Sea Ice. *Science* 339(6126): 1430-1432. doi: 10.1126/science.1231346.

Boyer, M, Aliaga, D, Pernov, JB, Angot, H, Quéléver, LLJ, Dada, L, Heutte, B, Dall'Osto, M, Beddows, DCS, Brasseur, Z, Beck, I, Bucci, S, Duetsch, M, Stohl, A, Laurila, T, Asmi, E, Massling, A, Thomas, DC, Nojgaard, JK, Chan, T, Sharma, S, Tunved, P, Krejci, R, Hansson, HC, Bianchi, F, Lehtipalo, K, Wiedensohler, A, Weinhold, K, Kulmala, M, Petäjä, T, Sipilä, M, Schmale, J, Jokinen, T. 2023. A full year of aerosol size distribution data from the central Arctic under an extreme positive Arctic Oscillation: insights from the Multidisciplinary drifting Observatory for the Study of Arctic Climate (MOSAiC) expedition. *Atmospheric Chemistry and Physics* 23(1): 389-415. doi: 10.5194/acp-23-389-2023

Butterworth, BJ, and Miller, S D 2016. Automated Underway Eddy Covariance System for Air–Sea Momentum, Heat, and $CO_2$ Fluxes in the Southern Ocean. *Journal of Atmospheric and Oceanic Technology,* 33, 635–652, doi:10.1175/JTECH-D-15-0156.1.

Cai, W-J, Chen, L, Chen, B, Gao, Z, Lee, SH, Chen, J, Pierrot, D, Sullivan, K, Wang, Y, Hu, X, Huang, W-J, Zhang, Y, Xu, S, Murata, A, Grebmeier, JM, Jones, EP, Zhang, H. 2010. Decrease in the CO2 Uptake Capacity in an Ice-Free Arctic Ocean Basin. *Science* 329:556-559. doi:10.1126/science.1189338

Campbell, K, Matero, I, Bellas, C, Turpin-Jelfs, T, Anhaus, P, Graeve, M, Fripiat, F, Tranter, M, Landy, JC, Sanchez-Baracaldo, P, Leu, E, Katlein, C, Mundy, CJ, Rysgaard, S, Tedesco, L, Haas, C, Nicolaus, M. 2022. Monitoring a changing Arctic: Recent advancements in the study of sea ice microbial communities. *Ambio* 51(2):318–332. doi:10.1007/s13280-021-01658-z.

Carmack, E, Polyakov, I, Padman, L, Fer, I, Hunke, E, Hutchings, J, Jackson, J, Kelley, D, Kwok, R, Layton, C, Melling, H, Perovich, D, Persson, O, Ruddick, B, Timmermans, ML, Toole, J, Ross, T, Vavrus, S, Winsor, P. 2015. Toward quantifying the increasing role of oceanic heat in sea ice loss in the new Arctic. *Bulletin of the American Meteorological Society* 96(12):2079–2105. doi:10.1175/bams-d-13-00177.1.

Chételat, J, McKinney, MA, Amyot, M, Dastoor, A, Douglas, TA, Heimbürger-Boavida, L-E, Kirk, J, Kahilainen, KK, Outridge, PM, Pelletier, N, Skov, H, St. Pierre, K, Vuorenmaa, J, Wang, F. 2022. Climate change and mercury in the Arctic: Abiotic interactions. *Science of the Total Environement* 824: 153715. doi:10.1016/j.scitotenv.2022.153715

Dada, L, Angot, H, Beck, I, Baccarini, A, Quéléver, LLJ, Boyer, M, Laurila, T, Brasseur, Z, Jozef, G, de Boer, G, Shupe, MD, Henning, S, Bucci, S, Dütsch, M, Stohl, A, Petäjä, T, Daellenbach, KR, Jokinen, T, Schmale, J. 2022. A central Arctic extreme aerosol event triggered by a warm air-mass intrusion. *Nature Communications* 13(1):5290. doi:10.1038/s41467-022-32872-2.

Dastoor, A, Angot, H, Bieser, J, Christensen, JH, Douglas, TA, Heimbürger-Boavida, LE, Jiskra, M, Mason, RP, McLagan, DS, Obrist, D, Outridge, PM, Petrova, MV, Ryjkov, A, St. Pierre, KA, Schartup, AT, Soerensen, AL, Toyota, K, Travnikov, O, Wilson, SJ, Zdanowicz, C. 2022. Arctic mercury cycling. *Nature Reviews Earth & Environment* 3(4):270–286. doi:10.1038/s43017-022-00269-w.

Deming, JW, Collins, ER. 2017. Sea ice as a habitat for bacteria, archaea and viruses. *Sea Ice*, D.N. Thomas (Ed.). Hoboken (NJ): Wiley. p. 326–351. doi:10.1002/9781118778371.ch13.

Dietz, R, Letcher, RJ, Aars, J, Andersen, M, Boltunov, A, Born, EW, Ciesielski, TM, Das, K, Dastnai, S, Derocher, AE, Desforges, P, Eulaers, I, Ferguson, S, Hallanger, IG, Heide-Jorgensen, MP, Heimbürger-Boavida, L-E, Hoekstra, PF, Jenssen, BM, Kohler, SG, Larsen, MM, Lindstrom, U, Lippold, A, Morris, A, Nabe-Nielsen, J, Nielsen, NH, Peacock, E, Pinzone, M, Rigét, FF, Rosing-Asvid, A, Routti, H, Siebert, U, Stenson, G, Stern, G, Strand, J, Sondergaard, J, Treu, G, Vikingsson, GA, Wang, F, Welker, JM, Wiig, O, Wilson, SJ, Sonne, C. 2022. A risk assessment review of mercury exposure in Arctic marine and terrestrial mammals. *Science of The Total Environment* 829: 154445. doi: doi:10.1016/j.scitotenv.2022.154445.

Ehrlich, J, Schaafsma, FL, Bluhm, BA, Peeken, I, Castellani, G, Brandt, A, Flores, H. 2020. Sympagic fauna in and under Arctic pack ice in the annual sea-ice system of the new Arctic. *Frontiers in Marine Science* 7:452. doi:10.3389/fmars.2020.00452.

Ekman, AM, Schmale, J. 2022. Aerosol processes in high-latitude environments and the effects on climate. In: *Aerosols and climate*. Amsterdam (NL): Elsevier. p. 651–706.

Else, BGT, Galley, RJ, Lansard, B, Barber, DG, Brown, KA, Miller, LA, Mucci, A, Papakyriakou, TN, Tremblay, J-É, and Rysgaard, S. 2013. Further observations of a decreasing atmospheric $CO_2$ uptake capacity in the Canada Basin (Arctic Ocean) due to sea ice loss, *Geophysical Research Letters* 40, doi:10.1002/grl.50268.

Falkowski, PG, Fenchel, T, Delong, EF. 2008. The microbial engines that drive Earth's biogeochemical cycles. *Science* 320(5879): 1034–1039. doi:10.1126/science.1153213.

Fernandez, RP, Berna, L, Tomazzeli, OG, Mahajan, AS, Li, Q, Kinnison, DE, Wang, S, Lamarque, JF, Tilmes, S, Skov, H, Cuevas, CA, Saiz-Lopez, A. 2024. Arctic halogens reduce ozone in the northern mid-latitudes. *Proceedings of the National Academy of Science USA* 121(38): e2401975121. doi:10.1073/pnas.2401975121.

Flores, H, Veyssière, G, Castellani, G, Wilkinson, J, Hoppmann, M, Karcher, M, Valcic, L, Cornils, A, Geoffroy, M, Nicolaus, M, Niehoff, B, Priou, P, Schmidt, K, Stroeve, J. 2023. Sea-ice decline could keep zooplankton deeper for longer. *Nature Climate Change* 13:1–9. doi:10.1038/s41558-023-01779-1.

Freitas, GP, Adachi, K, Conen, F, Heslin-Rees, D, Krejci, R, Tobo, Y, Yttri, KE, Zieger, P. 2023. Regionally sourced bioaerosols drive high-temperature ice nucleating particles in the Arctic. *Nature Communications* 14(1):5997. doi:10.1038/s41467-023-41696-7.

Frolov, IE, Gudkovich, ZM, Radionov, VF, Shirochkov, AV, Timokhov, LA. 2005. The Arctic Basin: Results from the Russian drifting stations. *Berlin (DE): Springer Praxis Books.* doi:10.1007/3-540-37665-8.

Fong, AA, Hoppe, CJM, Aberle, N, Ashjian, CJ, Assmy, P, Bai, Y, Bakker, DCE, Balmonte, JP, Barry, KR, Bertilsson, S, Boulton, W, Bowman, J, Bozzato, D, Bratbak, G, Buck, M, Campbell, RG, Castellani, G, Chamberlain, EJ, Chen, J, Chierici, M, Cornils, A, Creamean, JM, Damm, E, Dethloff, K, Droste, ES, Ebenhöh, O, Eggers, SL, Engel, A, Flores, H, Fransson, A, Frickenhaus, S, Gardner, J, Gelfman, CE, Granskog, MA, Graeve, M, Havermans, C, Heuzé, C, Hildebrandt, N, Hill, TCJ, Hoppema, M, Immerz, A, Jin, H, Koch, BP, Kong, X, Kraberg, A, Lan, M, Lange, BA, Larsen, A, Lebreton, B, Leu, E, Loose, B, Maslowski, W, Mavis, C, Metfies, K, Mock, T, Müller, O, Nicolaus, M, Niehoff, B, Nomura, D, Nöthig, EM, Oggier, M, Oldenburg, E, Olsen, LM, Peeken, I, Perovich, DK, Popa, O, Rabe, B, Ren, J, Rex, M, Rinke, A, Rokitta, S, Rost, B, Sakinan, S, Salganik, E, Schaafsma, FL, Schäfer, H, Schmidt, K, Shoemaker, KM, Shupe, MD, Snoeijs-Leijonmalm, P, Stefels, J, Svenson, A, Tao, R, Torres-Valdés, S, Torstensson, A, Toseland, A, Ulfsbo, A, Van Leeuwe, MA, Vortkamp, M, Webb, AL, Zhuang, Y, Gradinger, RR. 2024. Overview of the MOSAiC expedition: Ecosystem. *Elementa : Science of Anthropocene* 12(1):00135. doi:10.1525/elementa.2023.00135.

Gallo, F, Fossi, C, Weber, R, Santillo, D, Sousa, J, Ingram, I, Nadal, A, Romano, D. 2018. Marine litter plastics and microplastics and their toxic chemicals components: the need for urgent preventive measures. *Environmental Sciences Europe* 30(1): 13. doi: 10.1186/s12302-018-0139-z.

Gascard, J, Festy, J, le Goff, H, Weber, M, Bruemmer, B, Offermann, M, Doble, M, Wadhams, P, Forsberg, R, Hanson, S, Skourup, H, Gerland, S, Nicolaus, M, Metaxian, J, Grangeon, J, Haapala, J, Rinne, E, Haas, C, Wegener, A, Heygster, G, Jakobson, E, Palo, T, Wilkinson, J, Kaleschke, L, Claffey, K, Elder, B, Bottenheim, J. 2008. Exploring Arctic Transpolar Drift during dramatic sea ice retreat. *Eos Transactions American Geophysical Union* 89(3):21–22. doi:10.1029/2008eo030001.

Goosse, H, Kay, JE, Armour, KC, Bodas-Salcedo, A, Chepfer, H, Docquier, D, Jonko, A, Kushner, PJ, Lecomte, O, Massonnet, F, Park, HS, Pithan, F, Svensson, G, Vancoppenolle, M. 2018. Quantifying climate feedback in polar regions. *Nature Communications* 9(1):1919. doi:10.1038/s41467-018-04173-0.

Granskog, MA, Fer, I, Rinke, A, Steen, H. 2018. Atmosphere-ice-ocean-ecosystem processes in a thinner Arctic sea ice regime: The Norwegian Young Sea ICE (N-ICE2015) expedition. *Journal of Geophysical Research: Oceans* 123(3):1586–1594. doi:10.1002/2017jc013328.

Grosfeld, K, Treffeisen, R, Asseng, J, Bartsch, A, Bräuer, B, Fritzsch, B, Gerdes, R, Hendricks, S, Hiller, W, Heygster, G, Krumpen, T, Lemke, P, Melsheimer, C, Nicolaus, M, Ricker, R, Weigelt, M. 2016. Online sea-ice knowledge and data platform <www.meereisportal.de>. *Polarforschung*. doi:10.2312/polfor.2016.011.

Gupta, M, Marshall, J, Song, H, Campin, J, Meneghello, G. 2020. Sea-ice melt driven by ice-ocean stresses on the mesoscale. *Journal of Geophysical Research: Oceans* 125(11):e2020JC016404. doi:10.1029/2020JC016404.

Haine, TWN, Curry, B, Gerdes, R, Hansen, E, Karcher, M, Lee, C, Rudels, B, Spreen, G, de Steur, L, Stewart, KD, Woodgate, R. 2015. Arctic freshwater export: Status, mechanisms, and prospects. *Global and Planetary Change* 125:13–35. doi:10.1016/j.gloplacha.2014.11.013.

Heimbürger, L-E, Sonke, JE, Cossa, D, Point, D, Lagane, C, Laffont, L, Galfond, BT, Nicolaus, M, Rabe, B, van der Loeff, MR. 2015. Shallow methylmercury production in the marginal sea ice zone of the central Arctic Ocean. *Scientific Reports* 5:10318. doi:10.1038/srep10318.

Hermabessiere, L, Dehaut, A, Paul-Pont, I, Lacroix, C, Jezequel, R, Soudant, P, Duflos, G. 2017. Occurrence and effects of plastic additives on marine environments and organisms: A review. *Chemosphere* 182: 781-793. doi:10.1016/j.chemosphere.2017.05.096.

Heslin-Rees, D, Tunved, P, Ström, J, Cremer, R, Zieger, P, Riipinen, I, Ekman, AML, Eleftheriadis, K, Krejci, R. 2024. Increase in precipitation scavenging contributes to long-term reductions of light-absorbing aerosol in the Arctic. *Atmospheric Chemistry and Physics* 24(4): 2059-2075. doi: 10.5194/acp-24-2059-2024

Hop, H, Vihtakari, M, Bluhm, BA, Assmy, P, Poulin, M, Gradinger, R, Peeken, I, von Quillfeldt, C, Olsen, LM, Zhitina, L, Melnikov, IA. 2020. Changes in sea-ice protist diversity with declining sea ice in the Arctic Ocean from the 1980s to 2010s. *Frontiers in Marine Science* 7:243. doi:10.3389/fmars.2020.00243

Hoppe, CJM, Fuchs, N, Notz, D, Anderson, P, Assmy, P, Berge, J, Bratbak, G, Guillou, G, Kraberg, A, Larsen, A, Lebreton, B, Leu, E, Lucassen, M, Müller, O, Oziel, L, Rost, B, Schartmüller, B, Torstensson, A, Wloka, J. 2024. Photosynthetic light requirement near the theoretical minimum detected in Arctic microalgae. *Nature Communications* 15(1): 7385. doi: 10.1038/s41467-024-51636-8

Ibarbalz, FM, Henry, N, Mahé, F, Ardyna, M, Zingone, A, Scalco, E, Lovejoy, C, Lombard, F, Jaillon, O, Iudicone, D, Malviya, S, Tara Oceans Coordinators, Sullivan, MB, Chaffron, S, Karsenti, E, Babin, M, Boss, E, Wincker, P, Zinger, L, de Vargas, C, Bowler, C, Karp-Boss, L. 2023. Pan-Arctic plankton community structure and its global connectivity. *Elementa: Science of the Anthropocene* 11(1): 00060. doi:10.1525/elementa.2022.00060.

Jahn, A, Holland, MM, Kay, JE. 2024. Projections of an ice-free Arctic Ocean. *Nature Reviews Earth & Environment* 5(3):164–176. doi:10.1038/s43017-023-00515-9.

Kanhai, LDK, Gardfeldt, K, Krumpen, T, Thompson, RC, O'Connor, I. 2020. Microplastics in sea ice and seawater beneath ice floes from the Arctic Ocean. *Scientific Reports* 10(1):5004. doi:10.1038/s41598-020-61948-6.

Kay, JE, Gettelman, A. 2009. Cloud influence on and response to seasonal Arctic sea ice loss. *Journal of Geophysical Research: Atmospheres* 114(D18). doi:10.1029/2009jd011773.

Kay, JE, L'Ecuyer, T, Chepfer, H, Loeb, N, Morrison, A, Cesana, G. 2016. Recent advances in Arctic cloud and climate research. *Current Climate Change Reports* 2:159–169

Kohler, SG, Heimbürger-Boavida, L-E, Petrova, MV, Digernes, MG, Sanchez, N, Dufour, A, Simić, A, Ndungu, K, Ardelan, MV. 2022. Arctic Ocean's wintertime mercury concentrations limited by seasonal loss on the shelf. *Nature Geoscience* 15(8):621–626. doi:10.1038/s41561-022-00986-3.

Kohler, SG, Heimbürger-Boavida, L-E, Assmy, P, Müller, O, Thiele, S, Digernes, MG, Ndungu, K, Ardelan, MV. 2024. Biotic transformation of methylmercury at the onset of the Arctic spring bloom. *Progress in Oceanography* 222: 103224. doi:10.1016/j.pocean.2024.103224.

Krause, JW, Schulz, IK, Rowe, KA, Dobbins, W, Winding, MHS, Sejr, MK, Duarte, CM, Agusti, S. 2019. Silicic acid limitation drives bloom termination and potential carbon sequestration in an Arctic bloom. *Scientific Reports* 9(1):8149. doi:10.1038/s41598-019-44587-4

Krumpen, T, Belter, HJ, Boetius, A, Damm, E, Haas, C, Hendricks, S, Nicolaus, M, Nöthig, E-M, Paul, S, Peeken, I, Ricker, R, Stein, R. 2019. Arctic warming interrupts the Transpolar Drift and affects long-range transport of sea ice and ice-rafted matter. *Scientific Reports* 9:5459. doi:10.1038/s41598-019-41456-y

Krumpen, T, von Albedyll, L, Bünger, HJ, Castellani, G, Hartmann, J, Helm, V, Hendricks, S, Hutter, N, Landy, JC, Lisovski, S, Lüpkes, C, Rohde, J, Suhrhoff, M, Haas, C. 2025. Smoother sea ice with fewer pressure ridges in a more dynamic Arctic. *Nature Climate Change* 15(1): 66-72. doi: 10.1038/s41558-024-02199-5

Kvernvik, AC, Rokitta, SD, Leu, E, Harms, L, Gabrielsen, TM, Rost, B, Hoppe, CJM. 2020. Higher sensitivity towards light stress and ocean acidification in an Arctic sea-ice-associated diatom compared to a pelagic diatom. *New Phytologist* 226(6):1708–1724. doi:10.1111/nph.16501

Kwok, R. 2018. Arctic sea ice thickness, volume, and multiyear ice coverage: losses and coupled variability (1958–2018). *Environmental Research Letters* 13(10):105005. doi:10.1088/1748-9326/aae3ec

Landrigan, PJ, Raps, H, Cropper, M, Bald, C, Brunner, M, Canonizado, EM, Charles, D, Chiles, TC, Donohue, MJ, Enck, J, Fenichel, P, Fleming, LE, Ferrier-Pages, C, Fordham, R, Gozt, A, Griffin, C, Hahn, ME, Haryanto, B, Hixson, R, Ianelli, H, James, BD, Kumar, P, Laborde, A, Law, KL, Martin, K, Mu J, Mulders, Y, Mustapha, A, Niu, J, Pahl, S, Park, Y, Pedrotti, M-L, Pitt, JA, Ruchirawat, M, Seewoo, BJ, Spring, M, Stegeman, JJ, Suk, W, Symeonides, C, Takada, H, Thompson, RC, Vicini, A, Wang, Z, Whitman, E, Wirth, D, Wolff, M, Yousuf, AK, Dunlop, S. 2023. The Minderoo-Monaco Commission on Plastics and Human Health. *Annuals of Global Health* 89(1):23. doi:10.5334/aogh.4056

Law, KS, Roiger, A, Thomas, JL, Marelle, L, Raut, J-C, Dalsøren, S, Fuglestvedt, J, Tuccella, P, Weinzierl, B, Schlager, H. 2017. Local Arctic air pollution: Sources and impacts. *Ambio* 46(Suppl 3):453–463. doi:10.1007/s13280-017-0962-2

Law, KS, Hjorth, HL, Pernov, JB, Whaley, CH, Skov, H, Collaud, Coen, M, Langner, J, Arnold, SR, Tarasick, D, Christensen, J, Deushi, M, Effertz, P, Faluvegi, G, Gauss, M, Im, U, Oshima, N, Petropavlovskikh, I, Plummer, D, Tsigaridis, K, Tsyro, S, Solberg, S, Turnock, ST. 2023. Arctic Tropospheric Ozone Trends. *Geophysical Research Letters* 50(22): e2023GL103096. doi: https://doi.org/10.1029/2023GL103096

Lebreton, L, Andrady, A. 2019. Future scenarios of global plastic waste generation and disposal. *Palgrave Communications* 5(1): 6. doi: 10.1057/s41599-018-0212-7

Lebrun, M, Vancoppenolle, M, Madec, G, Massonnet, F. 2019. Arctic sea-ice-free season projected to extend into autumn. *Cryosphere* 13(1): 79-96. doi: 10.5194/tc-13-79-2019

Lenn, Y-D, Fer, I, Timmermans, M-L, MacKinnon, JA. 2022. Ocean mixing. In: *Encyclopedia of Ocean Sciences*, pp. 275–299. doi:10.1016/b978-0-12-821512-8.00018-9

Lewis, KM, van Dijken, GL, Arrigo, KR. 2020. Changes in phytoplankton concentration now drive increased Arctic Ocean primary production. *Science* 369(6500):198–202. doi:10.1126/science.aay8380

Lin, X, Massonnet, F, Fichefet, T, Vancoppenolle, M. 2023. Impact of atmospheric forcing uncertainties on Arctic and Antarctic sea ice simulations in CMIP6 OMIP models. *Cryosphere* 17(5):1935–1965. doi:10.5194/tc-17-1935-2023

Maillard, J, Ravetta, F, Raut, J-C, Mariage, V, Pelon, J. 2021. Characterisation and surface radiative impact of Arctic low clouds from the IAOOS field experiment. *Atmospheric Chemistry and Physics* 21(5):4079–4101. doi:10.5194/acp-21-4079-2021

Manucharyan, GE, Stewart, AL. 2022. Stirring of interior potential vorticity gradients as a formation mechanism for large subsurface-intensified eddies in the Beaufort Gyre. *Journal of Physical Oceanography* 52:3349–3370. doi:10.1175/jpo-d-22-0012.1

Falk, S, Reibig, L, Zilker, B, Richter, A, Sinnhuber, B-M. 2025. Challenges in simulating ozone depletion events in the Arctic boundary layer: A case study using ECHAM/MESSy for spring 2019/20. *EGUsphere* 2025:1–25. doi:10.5194/egusphere-2025-2.

Materić, D, Kjær, HA, Vallelonga, P, Tison, J-L, Röckmann, T, Holzinger, R. 2022. Nanoplastics measurements in Northern and Southern polar ice. *Environmental Research* 208:112741. doi:10.1016/j.envres.2022.112741

McCrystall, MR, Stroeve, J, Serreze, M, Forbes, BC, Screen, JA. 2021. New climate models reveal faster and larger increases in Arctic precipitation than previously projected. *Nature Communications* 12(1):6765. doi:10.1038/s41467-021-27031-y.

McKinney, MA, Chételat, J, Burke, SM, Elliott, KH, Fernie, KJ, Houde, M, Kahilainen, KK, Letcher, RJ, Morris, AD, Muir, DCG, Routti, H, Yurkowski, DJ. 2022. Climate change and mercury in the Arctic: Biotic interactions. *Science of the Total Environement* 834: 155221. doi: https://doi.org/10.1016/j.scitotenv.2022.155221.

Meneghello, G, Marshall, J, Lique, C, Isachsen, PE, Doddridge, E, Campin, J-M, Regan, H, Talandier, C. 2021. Genesis and decay of mesoscale baroclinic eddies in the seasonally ice-covered interior Arctic Ocean. *Journal of Physical Oceanography* 51(1): 115–129. doi: 10.1175/JPO-D-20-0054.1.

Mölders, N, Friberg, M. 2023. June to October aerosol optical depth over the Arctic at various spatial and temporal scales in MODIS, MAIAC, CALIOP and GOES data. *Open Journal of Air Pollution* 12(1):1–29. doi:10.4236/ojap.2023.121001.

Moline, MA, Karnovsky, NJ, Brown, Z, Divoky, GJ, Frazer, TK, Jacoby, CA, Torres, JJ, Fraser, WR. 2008. High latitude changes in ice dynamics and their impact on polar marine ecosystems. *Annals of the New York Academy of Sciences* 1134(1):267–319

Morison, J, Kwok, R, Dickinson, S, Andersen, R, Peralta-Ferriz, C, Morison, D, Rigor, I, Dewey, S, Guthrie, J. 2021. The cyclonic mode of Arctic Ocean circulation. *Journal of Physical Oceanography* 51(4):1053–1075. doi:10.1175/jpo-d-20-0190.1

Morris AD, Wilson, SJ, Fryer, RJ, Thomas, PJ, Hudelson, K, Andreasen, B, Blévin, P, Bustamante, P, Chastel, O, Christensen, G, Dietz, R, Evans, M, Evenset, A, Ferguson, SH, Fort, J, Gamberg, M, Grémillet, D, Houde, M, Letcher, RJ, Loseto, L, Muir, D, Pinzone, M, Poste, A, Routti, H, Sonne, C, Stern, G, Rigét, FF. 2022. Temporal trends of mercury in Arctic biota. *Science of the Total Environment* 848:157648. doi:10.1016/j.scitotenv.2022.157648.

Muilwijk, M, Smedsrud, LH, Ilicak, M, Drange, H. 2018. Atlantic Water heat transport variability in the 20th century Arctic Ocean from a global ocean model and observations. *Journal of Geophysical Research: Oceans* 123(11):8159–8179. doi:10.1029/2018jc014327

Muilwijk, M, Nummelin, A, Heuzé, C, Polyakov, IV, Zanowski, H, Smedsrud, LH. 2023. Divergence in Climate Model Projections of Future Arctic Atlantification. *Journal of Climate* 36(6): 1727-1748. doi: 10.1175/JCLI-D-22-0349.1

Muilwijk, M, Hattermann, T, Martin, T, Granskog, MA. 2024. Future sea ice weakening amplifies wind-driven trends in surface stress and Arctic Ocean spin-up. *Nature Communications* 15(1): 6889. doi: 10.1038/s41467-024-50874-0

Nicolaus, M, Katlein, C, Maslanik, J, Hendricks, S. 2012. Changes in Arctic sea ice result in increasing light transmittance and absorption. *Geophysical Research Letters.* doi:10.1029/2012GL053738.

Nicolaus, M, Perovich, DK, Spreen, G, Granskog, MA, von Albedyll, L, Angelopoulos, M, Anhaus, P, Arndt, S, Belter, HJ, Bessonov, V, Birnbaum, G, Brauchle, J, Calmer, R, Cardellach, E, Cheng, B, Clemens-Sewall, D, Dadic, R, Damm, E, de Boer, G, Demir, O, Dethloff, K, Divine, DV, Fong, AA, Fons, S, Frey, MM, Fuchs, N, Gabarro, C, Gerland, S, Goessling, HF, Gradinger, R, Haapala, J, Haas, C, Hamilton, J, Hannula, H-R, Hendricks, S, Herber, A, Heuzé, C, Hoppmann, M, Hoyland, KV, Huntemann, M, Hutchings, JK, Hwang, B, Itkin, P, Jacobi, H-W, Jaggi, M, Jutila, A, Kaleschke, L, Katlein, C, Kolabutin, N, Krampe, D, Kristensen, SS, Krumpen, T, Kurtz, N, Lampert, A, Lange, BA, Lei, R, Light,

B, Linhardt, F, Liston, GE, Loose, B, Macfarlane, AR, Mahmud, M, Matero, IO, Maus, S, Morgenstern, A, Naderpour, R, Nandan, V, Niubom, A, Oggier, M, Oppelt, N, Pätzold, F, Perron, C, Petrovsky, T, Pirazzini, R, Polashenski, C, Rabe, B, Raphael, IA, Regnery, J, Rex, M, Ricker, R, Riemann-Campe, K, Rinke, A, Rohde, J, Salganik, E, Scharien, RK, Schiller, M, Schneebeli, M, Semmling, M, Shimanchuk, E, Shupe, MD, Smith, MM, Smolyanitsky, V, Sokolov, V, Stanton, T, Stroeve, J, Thielke, L, Timofeeva, A, Tonboe, RT, Tavri, A, Tsamados, M, Wagner, DN, Watkins, D, Webster, M, Wendisch, M. 2022. Overview of the MOSAiC expedition: Snow and sea ice. *Elementa: Science of the Anthropocene* 10(1). doi:10.1525/elementa.2021.000046.

Niehaus, H, Spreen, G, Birnbaum, G, Istomina, L, Jäkel, E, Linhardt, F, Neckel, N, Fuchs, N, Nicolaus, M, Sperzel, T, Tao, R, Webster, M, Wright N. 2023. Sea ice melt pond fraction derived from Sentinel-2 data: Along the MOSAiC drift and Arctic-wide. *Geophysical Research Letters* 50(5). doi:10.1029/2022GL102102.

Nöthig, EM, Ramondenc, S, Haas, A, Hehemann, L, Walter, A, Bracher, A, Lalande, C, Metfies, K, Peeken, I, Bauerfeind, E, Boetius, A. 2020. Summertime chlorophyll a and particulate organic carbon standing stocks in surface waters of the Fram Strait and the Arctic Ocean (1991–2015). *Frontiers in Marine Science* 7:350. doi:10.3389/fmars.2020.00350.

Notz, D, SIMIP Community. 2020. Arctic Sea Ice in CMIP6. *Geophysical Research Letters* 47(10): e2019GL086749. doi:10.1029/2019GL086749.

Palmer, MA, Arrigo, KR, Mundy, CJ, Ehn, JK, Gosselin, M, Barber, DG, Martin, J, Alou, E, Roy, S, Tremblay, J-É. 2011. Spatial and temporal variation of photosynthetic parameters in natural phytoplankton assemblages in the Beaufort Sea, Canadian Arctic. *Polar Biology* 34(12):1915–1928. doi:10.1007/s00300-011-1050-x.

Oziel, L, Randelhoff, A, Ben, Jelloul, M. 2019. Borealization impacts shelf ecosystems across the Arctic. *Frontiers in Environmental Science* 7:1481420. doi:10.3389/fenvs.2019.1481420.

Perovich, DK, Andreas, EL, Curry, JA, Eiken, H, Fairall, CW, Grenfell, TC, Guest, PS, Intrieri, J, Kadko, D, Lindsay, RW, McPhee, MG, Morison, J, Moritz, RE, Paulson, CA, Pegau, WS, Persson, POG, Pinkel, R, Richter-Menge, JA, Stanton, T, Stern, H, Sturm, M, Tucker III, WB, Uttal, T. 1999. Year on ice gives climate insights. *Eos Transactions American Geophysical Union* 80(41):481–486. doi:10.1029/eo080i041p00481-01.

Persson, POG, Shupe, MD, Perovich, D, Solomon, A. 2017. Linking atmospheric synoptic transport, cloud phase, surface energy fluxes, and sea-ice growth: Observations of midwinter SHEBA conditions. *Climate Dynamics* 49:1341–1364.

Persson, O, Vihm,a T. 2017. The atmosphere over sea ice. In: *Sea Ice*. pp. 160–196.

Petäjä, T, Duplissy, E-M, Tabakova, K, Schmale, J, Altstädter, B, Ancellet, G, Arshinov, M, Balin, Y, Baltensperger, U, Bange, J, Beamish, A, Belan, B, Berchet, A, Bossi, R, Cairns, WRL, Ebinghaus, R, El Haddad, I, Ferreira-Araujo, B, Franck, A, Huang, L, Hyvärinen, A, Humbert, A, Kalogridis, A-C, Konstantinov, P, Lampert, A, MacLeod, M, Magand, O, Mahura, A, Marelle, L, Masloboev, V, Moisseev, D, Moschos, V, Neckel, N, Onishi, T, Osterwalder, S, Ovaska, A, Paasonen, P, Panchenko, M, Pankratov, F, Pernov, JB, Plastis, A, Popovicheva, O, Raut, J-C, Riandet, A, Sachs, T, Salvatori, R, Salzano, R, Schröder, L, Schön, Shevchenko, V, Skov, H, Sonke, JE, Spolaor, A, Stathopoulos, VK, Strahlendorff, M, Thomas, JL, Vitale, Vito, Vratolis, S, Barbante, C, Chabrillat, S, Dommergue, A, Eleftheriadis, K, Heilimo, J, Law, KS, Massling, A, Noe, SM, Paris, J-D, Prévôt, ASH, Riipinen, I, Wehner, B, Xie, Z, Lappalainen, H. 2020. Overview: Integrative and Comprehensive Understanding on Polar Environments (iCUPE) – concept and initial results. *Atmospheric Chemistry and Physics* 20(14):8551–8592.

Petrova, MV, Krisch, S, Lodeiro, P, Valk, O, Dufour, A, Rijkenberg, MJA, Achterberg, EP, Rabe, B, Rutgers van der Loeff, M, Hamelin, B, Sonke, JE, Garnier, C, Heimbürger-Boavida, L-É. 2020. Mercury species export from the Arctic to the Atlantic Ocean. *Marine Chemistry* 225: 103855. doi:10.1016/j.marchem.2020.103855.

Pistone, K, Eisenman, I, Ramanathan, V. 2019. Radiative heating of an ice-free Arctic Ocean. *Geophysical Research Letters* 46(13):7474–7480. doi:10.1029/2019gl082914

Pithan, F., and T. Mauritsen. 2014. Arctic amplification dominated by temperature feedbacks in contemporary climate models. *Nature Geoscience*, 7(3):181–184, doi:10.1038/ngeo2071.
Polyakov, IV, Timokhov, LA, Alexeev, VA, Bacon, S, Dmitrenko, IA, Fortier, L, Frolov, IE, Gascard, J-C, Hansen, E, Ivanov, VV, Laxon, S, Mauritzen, C, Perovich, D, Shimada, K, Simmons, HL, Sokolov, VT, Steele, M, Toole, J. 2010. Arctic Ocean warming contributes to reduced polar ice cap. *Journal of Physical Oceanography* 40(12):2743–2756. doi:10.1175/2010jpo4339.1.
Polyakov, IV, Pnyushkov, AV, Alkire, MB, Ashik, IM, Baumann, TM, Carmack, EC, Goszczko, I, Guthrie, J, Ivanov, VV, Kanzow T, Krishfield, R, Kwork, R, Sundfjord, A, Morison, J, Rember, R, Yulin, A. 2017. Greater role for Atlantic inflows on sea-ice loss in the Eurasian Basin of the Arctic Ocean. *Science* 356(6335):285–291. doi:10.1126/science.aai8204.
Polyakov, IV, Padman, L, Lenn, Y-D, Pnyushkov, A, Rember, R, Ivanov, VV. 2019. Eastern Arctic Ocean diapycnal heat fluxes through large double-diffusive steps. *Journal of Physical Oceanography* 49(1): 227–246. doi: 10.1175/JPO-D-18-0080.1
Polyakov, IV, Alkire, MB, Bluhm, BA, Brown, KA, Carmack, EC, Chierici, M, Slagstad, SL, Wassmann, P. 2020a. Borealization of the Arctic Ocean in response to anomalous advection from Sub-Arctic Seas. *Frontiers in Marine Science* 7:491. doi:10.3389/fmars.2020.00491.
Polyakov, IV, Rippeth, TP, Fer, I, Baumann, TM, Carmack, EC, Ivanov, VV, Janout, M, Padman, L, Pnyushkov, AV, Rember, R. 2020b. Intensification of near-surface currents and shear in the Eastern Arctic Ocean. *Geophysical Research Letters* 47(16). doi:10.1029/2020gl089469.
Polyakov, IV, Ingvaldsen, RB, Pnyushkov, AV, Bhatt, US, Francis, JA, Janout, M, Kwok, R, Skagseth, Ø. 2023 Fluctuating Atlantic inflows modulate Arctic atlantification. *Science (1979)* 381(6661): 972-979. doi: 10.1126/science.adh5158.
Polyakov, IV, Ballinger, TJ, Lader, R, Zhang, X. 2024. Modulated trends in Arctic surface air temperature extremes as a fingerprint of climate change. *Journal of Climate* 37(8):2381–2404. doi:10.1175/jcli-d-23-0266.1.
Porter, GCE, Adams, MP, Brooks, IM, Ickes, L, Karlsson, L, Leck, C, Salter, ME, Schmale, J, Siegel, K, Sikora, SNF, Tarn, MD, Vüllers, J, Wernli, H, Zieger, P, Zinke, J, Murray, BJ. 2022. Highly active ice-nucleating particles at the summer North Pole. *Journal of Geophysical Research: Atmospheres* 127(6):e2021JD036059. doi:10.1029/2021JD036059.
Proshutinsky, A, Krishfield, R, Timmermans, M-L. 2020. Introduction to Special Collection on Arctic Ocean Modeling and Observational Synthesis (FAMOS) 2: Beaufort Gyre Phenomenon. *Journal of Geophysical Research: Oceans* 125(2). doi:10.1029/2019jc015400.
Rabe, B, Karcher, M, Kauker, F, Schauer, U, Toole, JM, Krishfield, RA, Pisarev, S, Kikuchi, T, Su, J. 2014. Arctic Ocean basin liquid freshwater storage trend 1992–2012. *Geophysical Research Letters* 41(3):961–968. doi:10.1002/2013gl058121.
Rabe, B, Heuzé, C, Regnery, J, Aksenov, Y, Allerholt, J, Athanase, M, Bai, Y, Basque, C, Bauch, D, Baumann, TM, Chen, D, Cole, ST, Craw, L, Davies, A, Damm, E, Dethloff, K, Divine, DV, Doglioni, F, Ebert, F, Fang, Y-C, Fer, I, Fong, AA, Gradinger, R, Granskog, MA, Graupner, R, Haas, C, He, H, He, Y, Hoppmann, M, Janout, M, Kadko, D, Kanzow, T, Karam, S, Kawaguchi, Y, Koenig, Z, Kong, B, Krishfield, RA, Krumpen, T, Kuhlmey, D, Kuznetsov, I, Lan, M, Laukert, G, Lei, R, Li, T, Torres-Valdés, S, Lin, L, Lin, L, Liu, H, Liu, N, Loose, B, Ma, X, McKay, R, Mallet, M, Mallett, RDC, Maslowski, W, Mertens, C, Mohrholz, V, Muilwijk, M, Nicolaus, M, O'Brien, JK, Perovich, D, Ren, J, Rex, M, Ribeiro, N, Rinke, A, Schaffer, J, Schuffenhauer, I, Schulz, K, Shupe, MD, Shaw, W, Sokolov, V, Sommerfeld, A, Spreen, G, Stanton, T, Stephens, M, Su, J, Sukhikh, N, Sundfjord, A, Thomisch, K, Tippenhauer, S, Toole, JM, Vredenborg, M, Walter, M, Wang, H, Wang, L, Wang, Y, Wendisch, M, Zhao, J, Zhou, M, Zhu, J. 2022. Overview of the MOSAiC expedition: Physical oceanography. *Elementa: Science of the Anthropocene* 10(1). doi:10.1525/elementa.2021.00062.

Rabe, B, Cox, CJ, Fang, Y-C, Goessling, H, Granskog, MA, Hoppmann, M, Hutchings, JK, Krumpen, T, Kuznetsov, I, Lei, R, Li, T, Maslowski, W, Nicolaus, M, Perovich, D, Persson, O, Regnery, J, Rigor, I, Shupe, MD, Sokolov, V, Spreen, G, Stanton, T, Watkins, DM, Blockley, E, Buenger, HJ, Cole, S, Fong, A, Haapala, J, Heuzé, C, Hoppe, CJM, Janout, M, Jutila, A, Katlein, C, Krishfield, R, Lin, L, Ludwig, V, Morgenstern, A, O'Brien, J, Zurita, AQ, Rackow, T, Riemann-Campe, K, Rohde, J, Shaw, W, Smolyanitsky, V, Solomon, A, Sperling, A, Tao, R, Toole, J, Tsamados, M, Zhu, J, Zuo, G. 2024. The MOSAiC Distributed Network: Observing the coupled Arctic system with multidisciplinary, coordinated platforms. *Elementa: Science of the Anthropocene* 12(1). doi:10.1525/elementa.2023.00103.

Ramondenc, S, Nöthig, EM, Hufnagel, L, Bauerfeind, E, Busch, K, Knüppel, N, Kraft, A, Schröter, F, Seifert, M, Iversen, MH. 2023. Effects of Atlantification and changing sea-ice dynamics on zooplankton community structure and carbon flux between 2000 and 2016 in the eastern Fram Strait. *Limnology and Oceanography* 68:S39–S53. doi:10.1002/lno.12192.

Raut, J-C, Marelle, L, Fast, JD, Thomas, JL, Weinzierl, B, Law, KS, Berg, LK, Roiger, A, Easter, RC, Heimerl, K, Onishi, T, Delanoë, J, Schlager, H. 2017. Cross-polar transport and scavenging of Siberian aerosols containing black carbon during the 2012 ACCESS summer campaign. *Atmospheric Chemistry and Physics* 17(18):10969–10995. doi:10.5194/acp-17-10969-2017.

Raut, J-C, Law, KS, Onishi, T, Daskalakis, N, Marelle, L. 2022. Impact of shipping emissions on air pollution and pollutant deposition over the Barents Sea. *Environmental Pollution* 298:118832. doi:10.1016/j.envpol.2022.118832.

Raut, J-C, Thomas, JL, Marelle, L, Lapere, R. 2025. Sea ice, aerosols and clouds. In: Thomas DN, ed. *Sea Ice*. Hoboken, NJ: Wiley. doi:10.1002/9781394213764.ch18.

Randelhoff, A, Holding, J, Janout, M, Sejr, MK, Babin, M, Tremblay, J-É, Alkire, M. 2020. Pan-Arctic Ocean primary production constrained by turbulent nitrate fluxes. *Frontiers in Marine Science* 7. doi:10.3389/fmars.2020.00150.

Rantanen, M, Karpechko, AY, Lipponen, A, Nordling, K, Hyvärinen, O, Ruosteenoja, K, Vihma, T, Laaksonen, A. 2022. The Arctic has warmed nearly four times faster than the globe since 1979. *Communications Earth & Environment* 3(1):168. doi:10.1038/s43247-022-00498-3.

Rijkenberg, MJA, Slagter, HA, Rutgers van der Loeff, M, van Ooijen, J, Gerringa, LJA. 2018. Dissolved Fe in the deep and upper Arctic Ocean with a focus on Fe limitation in the Nansen Basin. *Frontiers in Marine Science* 5. doi:10.3389/fmars.2018.00088.

Rippeth, TP, Fine, EC. 2022. Turbulent mixing in a changing Arctic Ocean. *Oceanography* 35(1):60–73. doi:10.5670/oceanog.2022.103.

Rode, KD, Van Hemert, C, Wilson, RR, Woodruff, SP, Pabilonia, K, Ballweber, L, Kwok, O, Dubey, JP. 2024. Increased pathogen exposure of a marine apex predator over three decades. *PLoS ONE* 19(10):e0310973. doi:10.1371/journal.pone.0310973.

Runge, CA, Daigle, RM, Hausner, VH. 2020. Quantifying tourism booms and the increasing footprint in the Arctic with social media data. *PLoS ONE* 15(1):e0227189. doi:10.1371/journal.pone.0227189.

Schartup, AT, Soerensen, AL, Heimbürger-Boavida, L-E. 2020. Influence of the Arctic sea-ice regime shift on sea-ice methylated mercury trends. *Environmental Science and Technology Letters* 7: 708–713. doi:10.1021/acs.estlett.0c00465.

Schmale, J, Sharma, S, Decesari, S, Pernov, J, Massling, A, Hansson, H-C, von Salzen, K, Skov, H, Andrews, E, Quinn, PK, Upchurch, LM, Eleftheriadis, K, Traversi, R, Gilardoni, S, Mazzola, M, Laing, J, Hopke, P. 2022. Pan-Arctic seasonal cycles and long-term trends of aerosol properties from 10 observatories. *Atmospheric Chemistry and Physics* 22(5):3067–3096. doi:10.5194/acp-22-3067-2022.

Schmale, J, Flores, JM, Law, KS, Raut, J-C, O'Brien, J, Vardi, A, Koren, I, Ravetta, F, Bekki, S, Pazmino, A, Ardyna, M, Geoffroy, M, Lovejoy, C, Nicolaus, M, Babin, M, Bowler, C, Karp-Boss, L. 2025. Tara Polaris: Shedding light on microbial and climate feedback processes in the Arctic atmosphere. *Elementa: Science of the Anthropocene* 13(1): 00030. doi:10.1525/elementa.2025.00030.

Schweiger, AJ, Zhang, J, Lindsay, RW, Steele, M. 2008. Did unusually sunny skies help drive the record sea ice minimum of 2007? *Geophysical Research Letters* 35(10). doi:10.1029/2008GL033463.

Serreze, MC, Barry, RG. 2011. Processes and impacts of Arctic amplification: A research synthesis. *Global and Planetary Change* 77(1–2):85–96. doi:10.1016/j.gloplacha.2011.03.004.

Shupe, MD, Rex, M, Blomquist, B, Persson, POG, Schmale, J, Uttal, T, Althausen, D, Angot, H, Archer, S, Bariteau, L, Beck, I, Bilberry, J, Bucci, S, Buck, C, Boyer, M, Brasseur, Z, Brooks, IM, Calmer, R, Cassano, J, Castro, V, Chu, D, Costa, D, Cox, CJ, Creamean, J, Crewell, S, Dahlke, S, Damm, E, de Boer, G, Deckelmann, H, Dethloff, K, Dütsch, M, Ebell, K, Ehrlich, A, Ellis, J, Engelmann, R, Fong, AA, Frey, MM, Gallagher, MR, Ganzeveld, L, Gradinger, R, Graeser, J, Greenamyer, V, Griesche, H, Griffiths, S, Hamilton, J, Heinemann, G, Helmig, D, Herber, A, Heuzé, C, Hofer, J, Houchens, T, Howard, D, Inoue, J, Jacobi, H-W, Jaiser, R, Jokinen, T, Jourdan, O, Jozef, G, King, W, Kirchgaessner, A, Klingebiel, M, Krassovski, M, Krumpen, T, Lampert, A, Landing, W, Laurila, T, Lawrence, D, Lonardi, M, Loose, B, Lüpkes, C, Maahn, M, Macke, A, Maslowski, W, Marsay, C, Maturilli, M, Mech, M, Morris, S, Moser, M, Nicolaus, M, Ortega, P, Osborn, J, Pätzold, F, Perovich, DK, Petäjä, T, Pilz, C, Pirazzini, R, Posman, K, Powers, H, Pratt, KA, Preußer, A, Quéléver, L, Radenz, M, Rabe, B, Rinke, A, Sachs, T, Schulz, A, Siebert, H, Silva, T, Solomon, A, Sommerfeld, A, Stevens, R, Stohl, A, Svensson, G, Uin, J, Viegas, J, Voigt, C, von der Gathen, P, Wehner, B, Welker, JM, Wendisch, M, Werner, M, Xie, Z, Xu, J, Yildirim, E, Zhou, P. 2022. Overview of the MOSAiC expedition: Atmosphere. *Elementa: Science of the Anthropocene* 10(1):00060. doi:10.1525/elementa.2021.00060.

Smith, MM, Angot, H, Chamberlain, EJ, Droste, ES, Karam, S, Muilwijk, M, Webb, AL, Archer, SD, Beck, I, Blomquist, BW, Bowman, J, Boyer, M, Bozzato, D, Chierici, M, Creamean, J, D'Angelo, A, Delille, B, Fer, I, Fong, AA, Fransson, A, Fuchs, N, Gardner, J, Granskog, MA, Hoppe, CJM, Hoppema, M, Hoppmann, M, Mock, T, Muller, S, Müller, O, Nicolaus, M, Nomura, D, Petäjä, T, Salganik, E, Schmale, J, Schmidt, K, Schulz, KM, Shupe, MD, Stefels, J, Thielke, L, Tippenhauer, S, Ulfsbo, A, van Leeuwe, M, Webster, M, Yoshimura, M, Zhan, L. 2023. Thin and transient meltwater layers and false bottoms in the Arctic sea ice pack—Recent insights on these historically overlooked features. *Elementa: Science of the Anthropocene* 11(1):00025. doi:10.1525/elementa.2023.00025.

Sonne, C, Dietz, R, Jenssen, BM, Lam, SS, Letcher, RJ. 2021. Emerging contaminants and biological effects in Arctic wildlife. *Trends in Ecology & Evolution* 36(5): 421-429. Elsevier. doi: 10.1016/j.tree.2021.01.007.

Spall, M. 2008. Buoyancy-forced downwelling in boundary currents. *Journal of Physical Oceanography* 38: 2704–2721. doi:10.1175/2008JPO3993.1.

Spall, M. 2020. Potential vorticity dynamics of the Arctic halocline. *Journal of Physical Oceanography* 50(9): 2491–2506. doi:10.1175/jpo-d-20-0056.1.

Steele, M, Zhang, J, Ermold, W. 2010. Mechanisms of summertime upper Arctic Ocean warming and the effect on sea ice melt. *Journal of Geophysical Research: Oceans* 115(C11). doi:10.1029/2009jc005849.

Stroeve, J, Notz, D. 2018. Changing state of Arctic sea ice across all seasons. *Environmental Research Letters* 13(10):103001. doi:10.1088/1748-9326/aade56.

Sturm, M, Massom, RA. 2017. Sea Ice. In: Thomas DN, Dieckmann GS, eds. *Sea Ice*. Oxford: Wiley and Blackwell. pp. 65–109.

Sumata, H, de Steur, L, Divine, DV, Granskog, MA, Gerland, S. 2023. Regime shift in Arctic Ocean sea ice thickness. *Nature* 615(7952):443–449. doi:10.1038/s41586-022-05686-x.

Svensson, G, Mauritsen, T. 2020. Arctic Cloud Systems. In: *Clouds and Climate: Climate Science's Greatest Challenge*. Cambridge: Cambridge University Press. p. 297.

Tesan, J, Petrova, M, Puigcorbé, V, Black, EE, Valk, O, Dufour, A, Hamelin, B, Buesseler, K, Masqué, P, Le Moigne, FAC, Sonke, JE, van der Loeff, MR, Heimbürger-Boavida, L-E.

2020. Mercury export flux in the Arctic Ocean estimated from $^{234}$Th/$^{238}$U disequilibria. *ACS Earth and Space Chemistry* 4: 795–801. doi:10.1021/acsearthspacechem.0c00055.

Taylor, RL, Semeniuk, DM, Payne, CD, Zhou, J, Tremblay, J-É, Cullen, JT, Maldonado, MT. 2013. Colimitation by light, nitrate, and iron in the Beaufort Sea in late summer. *Journal of Geophysical Research: Oceans* 118:3260–3277. doi:10.1002/jgrc.20244.

Terhaar, J, Lauerwald, R, Regnier, P, Gruber, N, Bopp, L. 2021. Around one third of current Arctic Ocean primary production sustained by rivers and coastal erosion. *Nature Communications* 12(1):169. doi:10.1038/s41467-020-20470-z.

Timmermans, M-L, Marshall, J. 2020. Understanding Arctic Ocean circulation: A review of ocean dynamics in a changing climate. *Journal of Geophysical Research: Oceans* 125:e2018JC014378. doi:10.1029/2018JC014378.

Toole, JM, Timmermans, M-L, Perovich, DK, Krishfield, RA, Proshutinsky, A, Richter-Menge, JA. 2010. Influences of the ocean surface mixed layer and thermohaline stratification on Arctic sea ice in the central Canada Basin. *Journal of Geophysical Research: Oceans* 115:C10018. doi:10.1029/2009JC005660.

Tremblay, J-É, Gagnon, J. 2009. The effects of irradiance and nutrient supply on the productivity of Arctic waters: a perspective on climate change. Dordrecht: Springer Netherlands.

Tremblay, J-É, Anderson, LG, Matrai, P, Coupel, P, Bélanger, S, Michel, C, Reigstad, M. 2015. Global and regional drivers of nutrient supply, primary production and CO2 drawdown in the changing Arctic Ocean. *Progress in Oceanography* 139:171–196. doi:10.1016/j.pocean.2015.08.009.

Untersteiner, N. 1961. On the mass and heat budget of Arctic sea ice. *Archives of Meteorology, Geophysics and Bioclimatology Ser A* 12:151–182.

Uttal, T, Starkweather, S, Drummond, JR, Vihma, T, Makshtas, AP, Darby LS, Burkhart, JF, Cox, CJ, Schmeisser, LN, Haiden, T, Maturilli, M, Shupe, MD, De Boer, G, Saha, A, Grachev, AA, Crepinsek, SM, Bruhwiler, L, Goodison, B, McArthur, B, Walden, VP, Dlugokencky, EJ, Persson, POG, Lesins, G, Laurila, T, Ogren, JA, Stone, R, Long, CN, Sharma, S, Massling, A, Turner, DD, Stanitski, DM, Asmi, E, Aurela, M, Skov H, Eleftheriadis, K, Virkkula, A, Plattm Am Forland, EJ, Iijima, Y, Nielsen, IE, Bergin, MH, Candlish, L, Zimov, NS, Zimov, SA, O'Neill, NT, Fogal, PF, Kivi, R, Konopleva-Akish, EA, Verlinde, J, Kustov, VY, Vasel, B, Ivakhov, VM, Viisanen, Y, Intrieri, JM. 2016. International Arctic Systems for Observing the Atmosphere: An International Polar Year Legacy Consortium. *Bulletin of the American Meteorological Society* 97(6):1033–1056. doi:10.1175/BAMS-D-14-00045.1.

Vancoppenolle, M, Bopp, L, Madec, G, Dunne, J, Ilyina, T, Halloran, PR, Steiner, N. 2013. Future Arctic Ocean primary productivity from CMIP5 simulations: Uncertain outcome, but consistent mechanisms. *Global Biogeochemical Cycles* 27:1–15. doi:10.1002/gbc.20055.

Vancoppenolle, M, Deming, JW, Rysgaard, S, Fripiat, F, Acinas, SG, Babin, M, Ehn, J, Haapala, J, Hoppe, C, Karp-Boss, L, Kanaan, G, Maréchal, E, Mundy, CJ, Nicolaus, M, Søgaard, DH, Tedesco, L, Vincent, F. n.d. Microbial life in Arctic pack ice: Prospects from the Tara Polaris Expeditions. *Elementa: Science of the Anthropocene* (submitted).

Vihma, T. 2014. Effects of Arctic sea ice decline on weather and climate: A review. *Surveys in Geophysics* 35:1175–1214.

Vihma, T, Pirazzini, R, Fer, I, Renfrew, IA, Sedlar, J, Tjernström, M, Gascard, JC. 2014. Advances in understanding and parameterization of small-scale physical processes in the marine Arctic climate system: a review. *Atmospheric Chemistry and Physics* 14(17):9403–9450.

Von Albedyll, Lv, Hendricks, S, Grodofzig, R, Krumpen, T, Arndt, S, Belter, HJ, Birnbaum, G, Cheng, B, Hoppmann, M, Hutchings, J, Itkin, P, Lei, R, Nicolaus, M, Ricker, R, Rohde, J, Suhrhoff, M, Timofeeva, A, Watkins, D, Webster, M, Haas, C. 2022. Thermodynamic and dynamic contributions to seasonal Arctic sea ice thickness distributions from airborne observations. *Elementa: Science of the Anthropocene* 10:00074. doi:10.1525/elementa.2021.00074.

Von Friesen, LW, Riemann, L. 2020. Nitrogen fixation in a changing Arctic Ocean: An overlooked source of nitrogen? *Frontiers in Microbiology* 11: 596426. doi:10.3389/fmicb.2020.596426.

Waits, A, Emelyanova, A, Oksanen, A, Abass, K, Rautio, A. 2018. Human infectious diseases and the changing climate in the Arctic. *Environmental International* 121:703–713. doi:10.1016/j.envint.2018.09.042.

Wang, Y, Bi, H, Huang, H, Liu, Y, Liu, Y, Liang, X, Fu, M, Zhang, Z. 2019. Satellite-observed trends in the Arctic sea ice concentration for the period 1979–2016. *Journal of Oceanology and Limnology* 37(1):18–37. doi:10.1007/s00343-019-7284-0.

Wassmann, P, Reigstad, M. 2011. Future Arctic Ocean seasonal ice zones and implications for pelagic-benthic coupling. *Oceanography* 24(3): 220–231. doi:10.5670/oceanog.2011.74.

Webster, MA, Rigor, IG, Nghiem, SV, Kurtz, NT, Farrell, SL, Perovich, DK, Sturm, M. 2014. Interdecadal changes in snow depth on Arctic sea ice. *Journal of Geophysical Research: Oceans* 119(8):5395–5406. doi:10.1002/2014jc009985.

Webster, M, Holland, M, Wright, NC, Hendricks, S, Hutter, N, Itkin, P, Light, B, Linhardt, F, Perovich, DK, Raphael, IA, Smith, MM, von Albedyll, L, Zhang, J. 2022. Spatiotemporal evolution of melt ponds on Arctic sea ice. *Elementa: Science of the Anthropocene* 10:00072. doi:10.1525/elementa.2021.00072.

Woodgate, RA, Peralta-Ferriz, C. 2021. Warming and freshening of the Pacific inflow to the Arctic from 1990–2019 implying dramatic shoaling in Pacific Winter Water ventilation of the Arctic water column. *Geophysical Research Letters* 48:e2021GL092528. doi:10.1029/2021GL092528.

Xie, Z, Zhang, P, Wu, Z, Zhang, S, Wei, L, Mi L, Kuester, A, Gandrass, J, Ebinghaus, R, Yang, R, Wang, Z, Mi W. 2022. Legacy and emerging organic contaminants in the polar regions. *Science of The Total Environement* 835: 155376. doi:10.1016/j.scitotenv.2022.155376.

Yasunaka, S, Manizza, M, Terhaar, J, Olsen, A, Yamaguchi, R, Landschützer, P, Watanabe, E, Carroll, D, Adiwira, H, Müller, JD, Hauck J. 2023. An assessment of $CO_2$ uptake in the Arctic Ocean from 1985 to 2018. *Global Biogeochemical Cycles* 37: e2023GB007806. doi:10.1029/2023GB007806;

Yue, F, Angot, H, Blomquist, B, Schmale, J, Hoppe, CJM, Lei, R, Shupe, MD, Zhan, L, Ren, J, Liu, H, Beck, I, Howard, D, Jokinen, T, Laurila, T, Quéléver, L, Boyer, M, Petäjä, T, Archer, S, Bariteau, L, Helmig, D, Hueber, J, Jacobi, H-W, Posman, K, Xie, Z. 2023. The Marginal Ice Zone as a dominant source region of atmospheric mercury during central Arctic summertime. *Nature Communications* 14(1): 4887. doi: 10.1038/s41467-023-40660-9.

Zhang, Y, Yamamoto-Kawai, M, and Williams, WJ, 2020. Two decades of ocean acidification in the surface waters of the Beaufort Gyre, Arctic Ocean: effects of sea ice melt and retreat from 1997–2016. *Geophysical Research Letters* 47: e60119. doi: 10.1029/2019GL086421.

Zhuang, Y, Jin, H, Cai, W-J, Li, H, Qi, D, Chen, J. 2022. Extreme nitrate deficits in the western Arctic Ocean: Origin, decadal changes, and implications for denitrification on a polar marginal shelf. *Global Biogeochemical Cycles* 36(7): e2022GB007304. doi:10.1029/2022GB007304.

Zieger, P, Heslin-Rees, D, Karlsson, L, Koike, M, Modini, R, Krejci, R. 2023. Black carbon scavenging by low-level Arctic clouds. *Nature Communications* 14(1):5488. doi:10.1038/s41467-023-41221-w.

# Tables

Table 1. Sentinels of the changing Arctic Ocean and their roles in the climate and ecosystem, for study during Tara Polaris expeditions at Tara Polar Station.

| Sentinel[a] | Role |
|---|---|
| **Atmosphere** | |
| Meteorology | Air temperature, humidity and winds of different levels are key indicators of the climate system and Arctic warming. |
| Clouds | Clouds control radiative transfer and water vapor transport, thus playing a key role in Arctic Amplification. |
| Trace gases | Trace gas concentrations reflect biospheric metabolism and diverse human activities. |
| Aerosols | Aerosols are indicators of regional human activity: long-range/local natural and anthropogenic emissions. |
| Contaminants (mercury, microplastics, other chemicals of emerging Arctic concern) | Atmospheric concentrations and deposition of contaminants reflect long-range transport, seasonal dynamics, and chemical processing in the Arctic atmosphere. |
| **Sea ice and snow** | |
| Snow cover | Snow thickness and properties control surface properties, the surface energy budget and under-ice light. |
| Ice cover | Sea-ice thickness and physical properties control heat, mass, and momentum exchange between atmosphere and ocean. |
| Ice dynamics | Drift velocity, pressure ridges and leads are key elements for momentum transfer and shape sea-ice habitats. |
| Ice habitat and biodiversity | Ranging from microbes to mammals, sea ice is a habitat and key factor of the polar ecosystem. |
| Contaminants (mercury, microplastics, other chemicals of emerging Arctic concern) | Sea ice and snow act as temporary sinks and transport media for contaminants, influencing their redistribution and availability to ecosystems during melt. |
| **Ocean** | |
| Ocean hydrography | Temperature and salinity control the state of the ocean; freshwater is a key factor in ocean stratification. |
| Ocean dynamics | Dynamics and eddies play a key role in redistributing physical and biogeochemical properties. |
| Ocean mixing | Topography and winds cause ocean mixing that impacts ocean stratification, sea ice, and water mass properties. |
| Under-ice light | Changing sea-ice conditions increase light availability, and change in seasonal timing has strong impacts for the ecosystem. |
| Nutrient availability | Changing nutrient concentrations and ratios affect the productivity and structure of marine ecosystems. |
| Ocean biodiversity | Diversity of bacteria, plankton, fish, and marine mammals are indicators of changes in ocean, ice, and atmosphere. |
| Microbial functioning | Microbial ecosystem functioning reacts to environmental changes, and thus future functional changes are expected. |
| Carbon export | Quantity and quality of carbon exports into the ocean depend on biophysical drivers and impact food webs. |
| Contaminants (mercury, microplastics, other chemicals of emerging Arctic concern) | Long-term and seasonal trends of contaminants reflect ocean circulation, stratification, and biogeochemical processes, with implications for food web exposure and ecosystem health. |

[a]Categorized by compartment, as described in more detail in Section 3; some sentinels cross compartments.

**Table 2.** Main parameters of the long-term measurements at Tara Polar Station.

| Parameter[a] | Method[b] | Spatial scale (horizontal[c] / vertical[d]) | Time interval |
|---|---|---|---|
| **Atmosphere** | | | |
| Basic meteorology | Automatic weather station | On TPS / 10 m | Minutes |
| | Automatic weather station | On ice / 2–10 m | Minutes |
| Fluxes, heat and momentum | Meteorological tower | On ice / 2–10 m | Minutes |
| Atmospheric sounding | Balloon launches | On TPS / 0–700 m | Daily |
| Fluxes, radiation, integrated | Automatic weather station | On TPS / 3 m | Minutes |
| | Automatic weather station | On ice / 2–10 m | Minutes |
| Fluxes, radiation, spectral | Radiation station | On TPS / 10 m | Minutes |
| | Radiation station | On ice / 2 m | Minutes |
| | Transects across ice | 0–200 m / 1 m | Weekly |
| Cloud distribution (cover, height, optical depth, phase partitioning) | Radar | On TPS / 0.1–10 km | Minutes |
| Trace gases ($O_3$, CO, $CH_4$) and VOC | Analyzers | On TPS / 10 m | Weekly |
| | Sorbent tubes for BVOC | On TPS / 0–700 m | Weekly |
| $CO_2$ concentration | Infrared gas analyzer | On TPS / 10 m | Minutes |
| Aerosol microphysics, vertical distribution, cloud condensation nuclei | Lidar | On TPS / 0.1–10 km | Minutes |
| | Balloon launches | On TPS / 0–700 m | Minutes |
| Black carbon, organic carbon, isotopes, contaminants (microplastics, other chemicals of emerging Arctic concern) | Aethalometer, filter samplers and laboratory analysis | On TPS / 10 m | Minutes to months |
| Mercury | Automated gaseous mercury analyzer | On TPS / 10 m | Minutes |
| **Sea ice and snow** | | | |
| Sea ice thickness | Transects across ice | 1 km, each 1 m | Weekly |
| | ROV transect | 200 x 200 m, each 1 m | Weekly |
| | Autonomous station | On ice | Hours |
| Snow thickness | Transect across ice | 1 km, each 1 m | Weekly |
| | Autonomous station | On ice | Hours |
| Freeboard | Transect across ice | 100 m, each 10 m | Weekly |
| Sea-ice density | Coring, weight/volume | 1 station | Weekly |
| Snow density | Transect across ice | 100 m, each 10 m | Weekly |
| Sea-ice salinity | Coring, sampling | 1 station | Weekly |
| Sea-ice temperature | Autonomous station | On ice | Hours |
| | Coring | 1 station | Weekly |
| Snow salinity | Snow pit | 1 station | Weekly |
| Sea-ice texture | Coring, sampling | 1 station | Weekly |
| Snow stratigraphy | Transect across ice | 100 m, each 10 m | Weekly |
| Surface albedo (Apr–Oct) | Snow pit | 1 station | Weekly |
| | Transect across ice | 200 m, each 10 m | Weekly |
| | Autonomous station | On ice | Minutes |
| | Drone survey | 500 x 500 m, each 1 m | Weekly |
| Light transmittance | ROV transects | 200 x 200 m, each 1 m | Weekly |
| | Autonomous station | stationary | Minutes |
| Melt pond properties | Drone survey | 500 x 500 m, each 1 m | Weekly |
| Snow impurities | Snow pit, sampling | 1 station | Weekly |
| Roughness/topography | ROV transects | 200 x 200 m, each 1 m | Weekly |
| Sea ice biogeochemical and ecological parameters | Coring, sampling | 1 station | Weekly |
| Surface conditions (ice types, leads) | Drone survey | 500 x 500 m | Weekly |
| Contaminants (mercury, microplastics, other chemicals of emerging Arctic concern) | Coring, snow pit | 1 station, higher resolution during polar sunrise | Monthly |
| **Ocean** | | | |
| Temperature | CTD rosette | On TPS / 0–2000 m | Weekly |
| | ROV transects | 400 m / 0–50 m | Weekly |
| | Autonomous | On ice / 0–100m | Minutes |
| | In-line system (TSG) | On TPS / 2.5 m | Minutes |
| Salinity | CTD rosette | On TPS / 0–2000 m | Weekly |
| | ROV transects | 400 m / 0–50 m | Weekly |
| | Autonomous | On ice / 0–100 m | Minutes |
| | In-line system (TSG) | On TPS / 2.5 m | Minutes |
| Density | CTD rosette | On TPS / 0–2000 m | Weekly |
| Dissolved oxygen | CTD rosette | On TPS / 0–2000 m | Weekly |
| | In-line system | On TPS / 2.5 m | Minutes |
| pH | In-line system | On TPS / 2.5 m | Minutes |
| $p$CO$_2$ | In-line system | On TPS / 2.5 m | Minutes |
| Chlorophyll *a* fluorescence | CTD rosette | On TPS / 0–2000 m | Weekly |
| | ROV transects | 400 m / 0–50 m | Weekly |
| | In-line system | On TPS / 2.5 m | Minutes |
| Nitrate | CTD rosette | On TPS / 0–2000 m | Weekly |
| CDOM fluorescence | CTD rosette | On TPS / 0–2000 m | Weekly |
| Current velocity, 3D | ADCP | On ice | Seconds |
| Particulate matter, plankton | UVP on CTD rosette | On TPS / 0–2000 m | Weekly |
| | Sediment traps | On ice / 30 and 100 m | Weekly–monthly |
| Suspended particles | Imaging Flow Cyto Bot | On TPS / 0–40 m | Weekly |
| Bacteria, phytoplankton, zooplankton and fish communities and abundances | Water samples | On TPS / station ice | Weekly |
| | Nets | On TPS / station ice | Weekly |
| | Acoustic Backscatter | On TPS / 0–2500 m | Seconds |
| | UVP | On TPS / 0–2500 m | Weekly |
| Light in ocean | Autonomous station | On ice / 0–50 m | Minutes |
| Ambient noise, mammals | Hydrophone | On ice | Minutes |
| Bio-optical properties | Inline system | On TPS / 2.5 m | Minutes |
| Contaminants (mercury, microplastics and other chemicals of emerging | CTD rosette, nets | On TPS / 0–2000 m | Weekly–quarterly |

Arctic concern)

[a]Parameters obtained with different methods, representing different spatial scales and realized in different intervals, have multiple entries; abbreviated parameters are volatile organic carbon (VOC) and chromophoric dissolved organic matter (CDOM).,
[b]Biological VOC (BVOC), remotely operated vehicle (ROV), conductivity-temperature-depth (CTD),.acoustic doppler current profiler (ADCP), ultrasonic velocity profiler (UVP)
[c]On TPS indicates stationary measurements on the platform Tara Polar Station (TPS); on ice indicates stationary measurements on the ice.
[d]Vertical range for Atmosphere is above surface; for Sea ice and snow, it is not applicable (none specified); for Ocean, it is below surface.

Figures

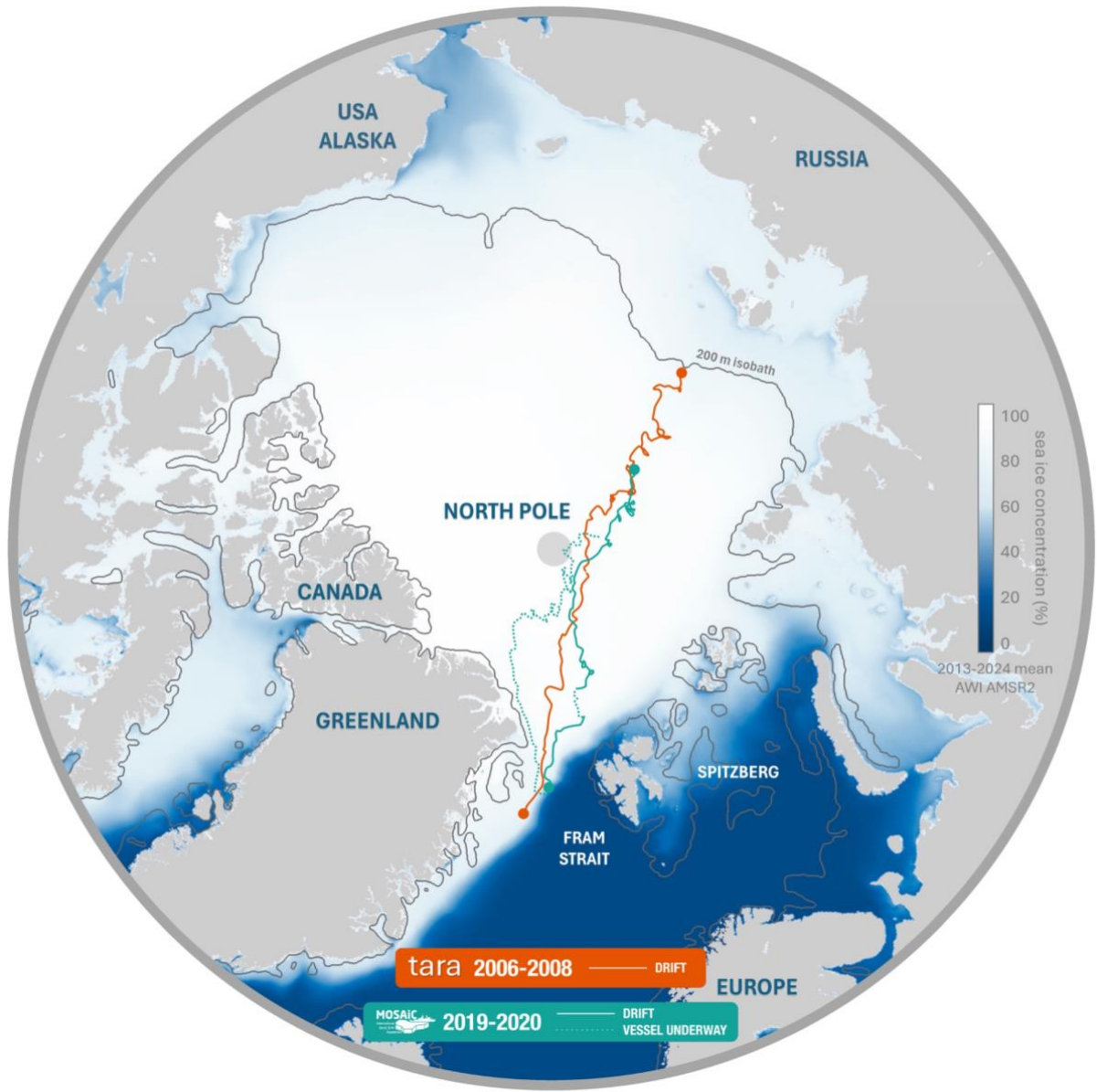

**Figure 1. Map of the central Arctic Ocean with drift trajectories of two recent drift experiments.** The Tara (orange) and MOSAiC (teal) drift experiments built upon prior experiments (see Section 1). Also shown are the mean sea-ice concentrations from 2013 to 2024 (color scale bar; from daily AWI AMSR2 ice concentration product v110) and the 200 m isobath (gray line; from ETOPO Global Relief Model).

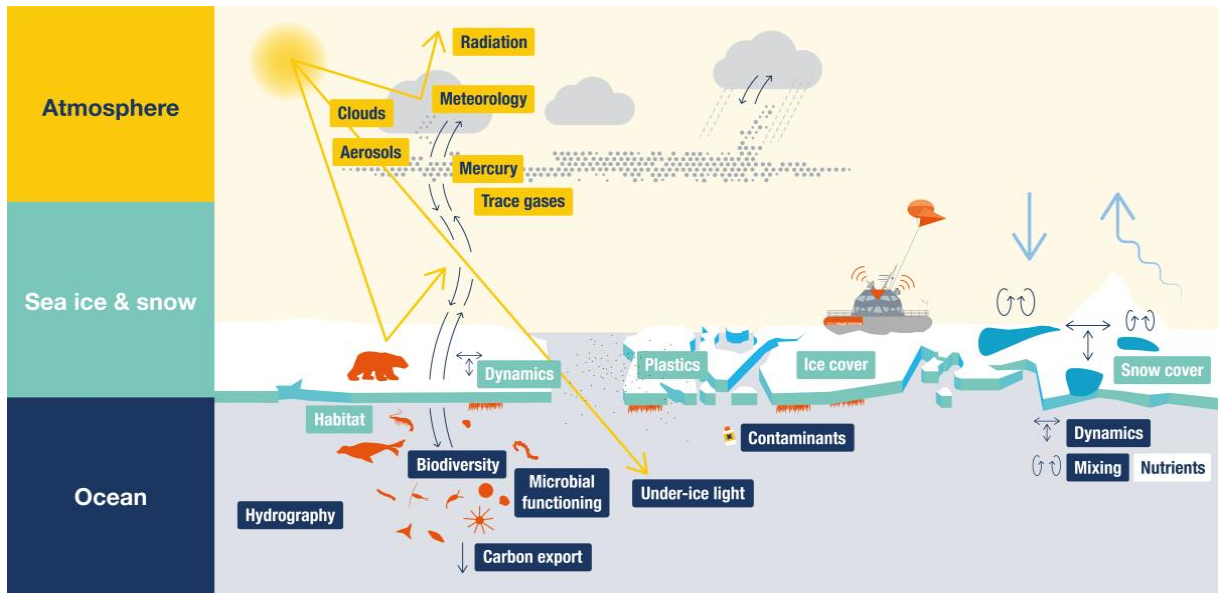

**Figure 2. The coupled system of the central Arctic Ocean.** Color-coded labels show the sentinels of the atmosphere (yellow), sea ice and snow (teal), and ocean (dark blue) under study by Tara Polar Station and described further in Table 1; biota and nutrients are also highlighted (orange).

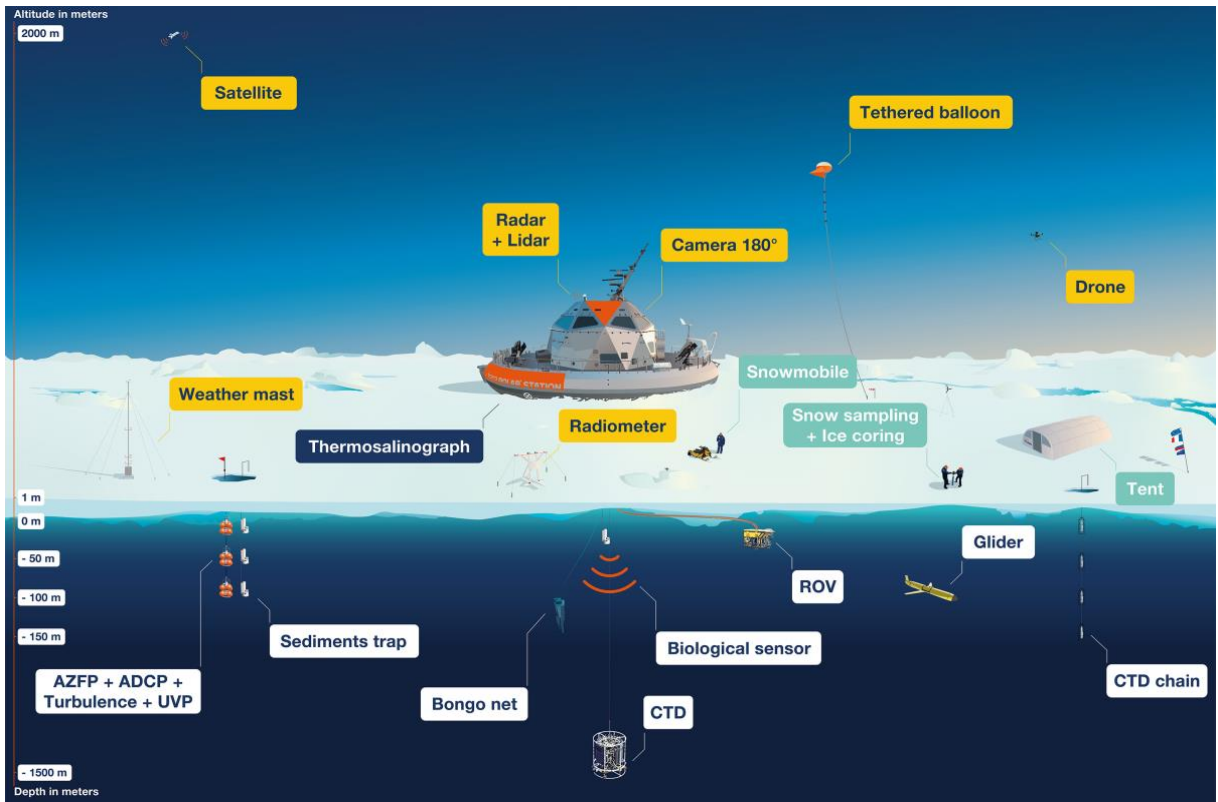

**Figure 3. Illustration of the main methodological approaches for long-term monitoring at the Tara Polar Station.** Methods are color-coded by atmosphere (yellow), sea ice and snow (teal), and ocean (white), including biological sensors (orange). Methods are described further in Section 4; key parameters are provided in Table 2. Abbreviations are for remotely operated vehicle (ROV), Acoustic Zooplankton and Fish Profiler (AZFP), acoustic Doppler current profiler (ADCP), and conductivity-temperature-depth (CTD).

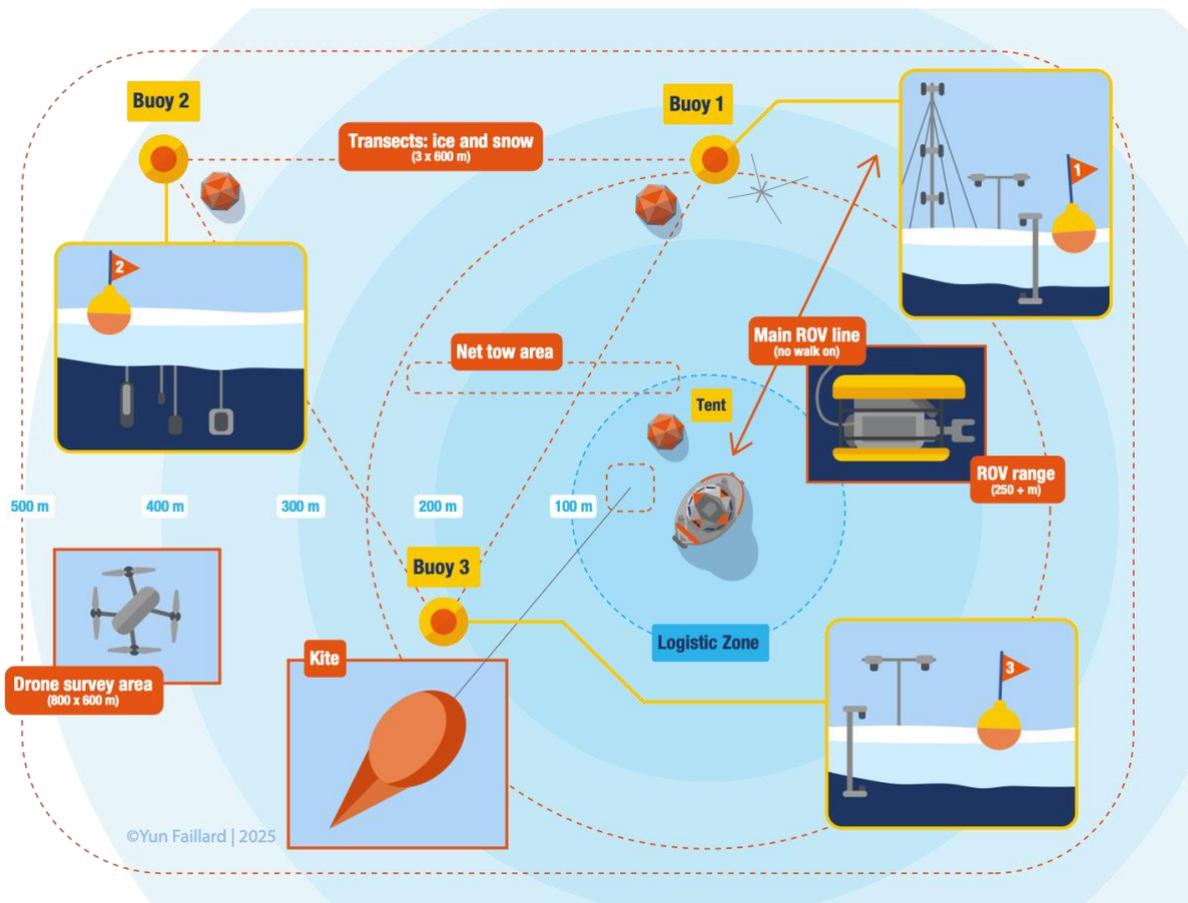

**Figure 4. Ice station layout in the vicinity of Tara Polar Station.** The illustration shows how the different observational components overlay and interact with each other. The observational concept is described in Section 4, and the main parameters and observational scales are given in Table 2. Abbreviations are for remotely operated vehicle (ROV).